\documentclass[preprint,
superscriptaddress,
nofootinbib,
 amsmath,amssymb,
 aps,
longbibliography
]{revtex4-2}

\usepackage{graphicx}
\usepackage{dcolumn}
\usepackage{bm}

\usepackage[normalem]{ulem}
\usepackage{nicefrac}
\usepackage{xcolor}
\usepackage{bm}
\usepackage{multirow}
\usepackage{amsfonts}
\usepackage{amssymb}
\usepackage{bbold}
\usepackage{hyperref}
\usepackage{capt-of}
\usepackage{float}
\usepackage{lipsum}
\usepackage{soul}

\usepackage{color}
\usepackage{tikz}
\usetikzlibrary{shapes}

\definecolor{minered}{HTML}{FF0000}
\definecolor{mineblue}{HTML}{4169E1}
\definecolor{minedarkblue}{HTML}{AFEEEE}
\definecolor{minegreen}{HTML}{228B22}
\definecolor{colorCa01}{HTML}{FA9070}
\definecolor{colorCa015}{HTML}{F37457}
\definecolor{colorCa02}{HTML}{EC583E}
\definecolor{colorCa03}{HTML}{CF2613}
\definecolor{colorCa04}{HTML}{7A180C}
\definecolor{colorCa045}{HTML}{511008}
\definecolor{colorCa055}{HTML}{000000}

\newcommand{\circlea}{\raisebox{0.5pt}{\tikz{\node[draw,scale=0.5,circle,fill=colorCa01](){};}}}
\newcommand{\circleb}{\raisebox{0.5pt}{\tikz{\node[draw,scale=0.5,circle,fill=colorCa02](){};}}}
\newcommand{\circlec}{\raisebox{0.5pt}{\tikz{\node[draw,scale=0.5,circle,fill=colorCa03](){};}}}
\newcommand{\circled}{\raisebox{0.5pt}{\tikz{\node[draw,scale=0.5,circle,fill=colorCa04](){};}}}

\newcommand{\circleBLUE}{\raisebox{0.5pt}{\tikz{\node[draw,scale=0.5,circle,fill=mineblue](){};}}}

\newcommand{\rhombusBLUE}{\raisebox{0.5pt}{\tikz{\node[draw,scale=0.4,regular polygon,regular polygon sides=4,fill=minedarkblue, rotate=45](){};}}}

\newcommand{\pentagonBLUE}{\raisebox{0.5pt}{\tikz{\node[draw,scale=0.5,regular polygon,regular polygon sides=5,fill=minedarkblue, rotate=0](){};}}}

\newcommand{\trianglea}{\raisebox{0.5pt}{\tikz{\node[draw,scale=0.3,regular polygon, regular polygon sides=3,fill=colorCa01,rotate=0](){};}}}

\newcommand{\triangleb}{\raisebox{0.5pt}{\tikz{\node[draw,scale=0.3,regular polygon, regular polygon sides=3,fill=colorCa02,rotate=0](){};}}}

\newcommand{\trianglec}{\raisebox{0.5pt}{\tikz{\node[draw,scale=0.3,regular polygon, regular polygon sides=3,fill=colorCa03,rotate=0](){};}}}

\newcommand{\triangled}{\raisebox{0.5pt}{\tikz{\node[draw,scale=0.3,regular polygon, regular polygon sides=3,fill=colorCa04,rotate=0](){};}}}

\newcommand{\triangleRED}{\raisebox{0.5pt}{\tikz{\node[draw,scale=0.3,regular polygon,regular polygon sides=3,fill=minered,rotate=0](){};}}}

\newcommand{\squarea}{\raisebox{0.5pt}{\tikz{\node[draw,scale=0.4,regular polygon, regular polygon sides=4,fill=colorCa015](){};}}}
\newcommand{\squareb}{\raisebox{0.5pt}{\tikz{\node[draw,scale=0.4,regular polygon, regular polygon sides=4,fill=colorCa03](){};}}}
\newcommand{\squarec}{\raisebox{0.5pt}{\tikz{\node[draw,scale=0.4,regular polygon, regular polygon sides=4,fill=colorCa045](){};}}}
\newcommand{\squared}{\raisebox{0.5pt}{\tikz{\node[draw,scale=0.4,regular polygon, regular polygon sides=4,fill=colorCa055](){};}}}
\newcommand{\squareGREEN}{\raisebox{0.5pt}{\tikz{\node[draw,scale=0.4,regular polygon, regular polygon sides=4,fill=minegreen](){};}}}

\newcommand{\Ca}{\text{Ca}}
\newcommand{\Rey}{\text{Re}}
\newcommand{\Uw}{\vec{U}_{\mbox{\scriptsize w}}}

\newcommand{\Cacr}{\text{Ca}_{\mbox{\scriptsize cr}}}

\newcommand{\effi}{\mbox{n}_i}

\newcommand{\cs}{c_{\mbox{\scriptsize s}}}

\newcommand{\Lx}{L_{\mbox {\tiny x}}}
\newcommand{\Ly}{L_{\mbox {\tiny y}}}
\newcommand{\Lz}{L_{\mbox {\tiny z}}}

\newcommand{\gammadot}{\dot{\gamma}}

\newcommand{\trace}[1]{\mbox{tr}\left(#1\right)}

\renewcommand{\vec}[1]{\boldsymbol{\bm{#1}}} 

\newcommand{\fmmone}{f_1^{(\mbox{\tiny MM})}}
\newcommand{\fmmtwo}{f_2^{(\mbox{\tiny MM})}}

\newcommand{\fmmonetwo}{f_{1,2}^{(\mbox{\tiny MM})}}

\begin{document}

\preprint{APS/123-QED}

\title{A reduced model for droplet dynamics in shear flows \\ at finite capillary numbers}
\author{Diego Taglienti}
\email{diego.taglienti@roma2.infn.it}
\affiliation{%
Department of Physics \& INFN, University of Rome ``Tor Vergata'', Via della Ricerca Scientifica 1, 00133, Rome, Italy.
}%
\author{Fabio Guglietta}%
\affiliation{Helmholtz Institute Erlangen-Nürnberg for Renewable Energy (IEK-11), Forschungszentrum Jülich GmbH, Cauerstraße 1, 91058 Erlangen, Germany}
\author{Mauro Sbragaglia}
\affiliation{%
Department of Physics \& INFN, University of Rome ``Tor Vergata'', Via della Ricerca Scientifica 1, 00133, Rome, Italy.
}%


\date{\today}

\begin{abstract}
We propose an extension of the Maffettone-Minale (MM) model to predict droplet dynamics in shear flow. The parameters of the MM model are traditionally retrieved in the framework of the perturbation theory for small deformations, i.e., small capillary numbers ($\Ca \ll 1$) applied to Stokes equations.
In this work, we take a novel route, in that we determine the model parameters at finite capillary numbers ($\Ca\sim {\cal O}(1)$) without relying on perturbation theory results, while retaining a realistic representation in loading time and steady deformation attained by the droplet for different realizations of the viscosity ratio $\lambda$ between the inner and the outer fluids. This extended MM (EMM) model hinges on an independent characterization of the process of droplet deformation via fully three-dimensional numerical simulations of Stokes equations employing the Immersed Boundary - Lattice Boltzmann (IB-LB) numerical techniques. Issues on droplet breakup are also addressed and discussed within the EMM model.
\end{abstract}

\maketitle


\section{\label{sec:intro}Introduction}

\begin{figure*}
\centering
\includegraphics[width=0.75\linewidth]{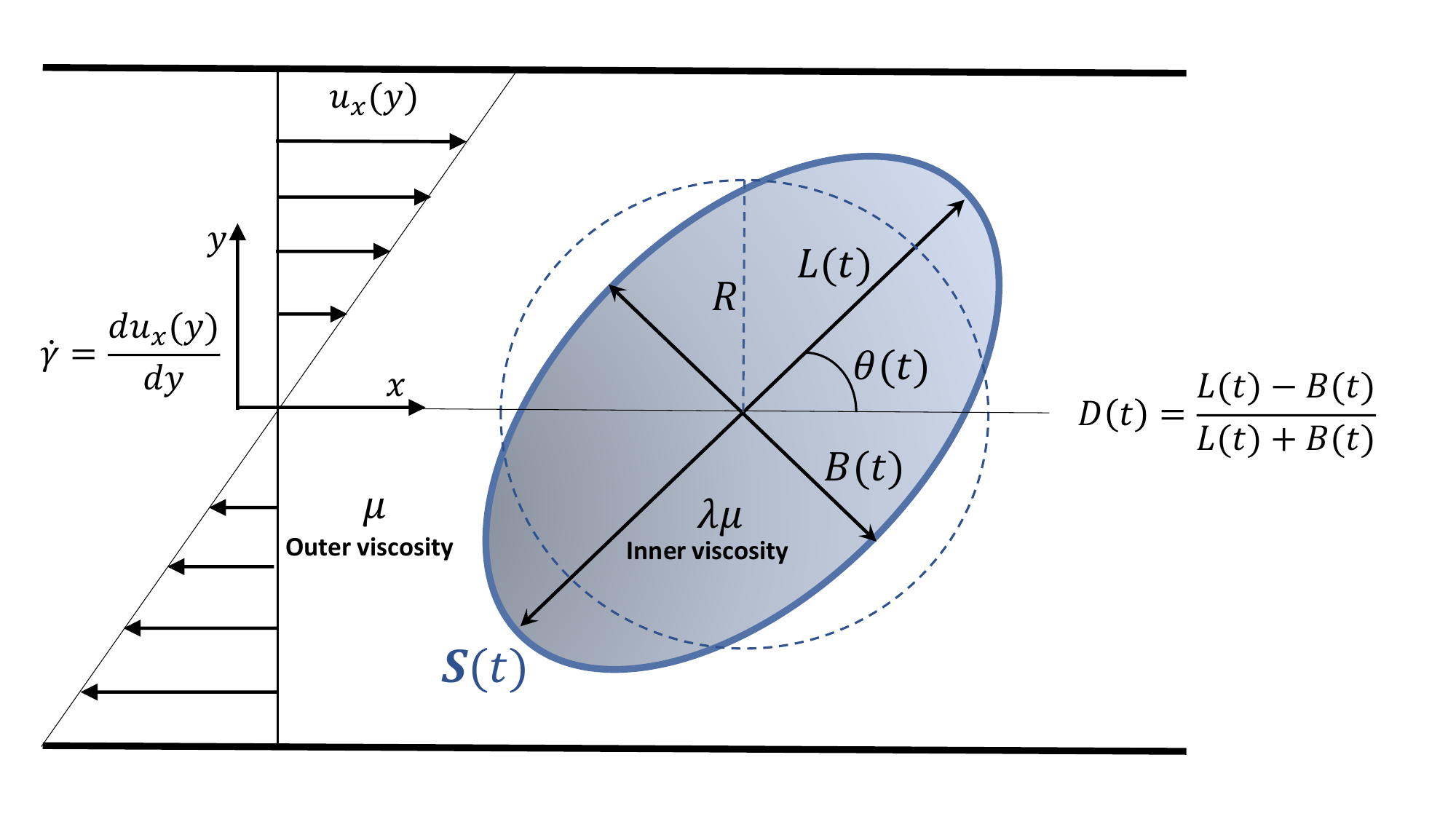}%
\caption{Shear plane section of a droplet with viscosity $ \lambda \mu$ deforming under the effect of a simple shear flow with intensity $\gammadot$ applied in an outer fluid with visocisty $\mu$. The droplet has an initial spherical shape with radius $R$ and deforms into a non-spherical shape due to the action of the shear flow. The shape of the droplet is described via a second-order tensor $\vec{S}(t)$. The major ($L(t)$) and minor ($B(t)$) axes in the shear plane are used to compute the time dependent Taylor index $D(t)$, while $\theta(t)$ measures the inclination angle with respect to the streamflow direction ($x-$axis).}\label{fig:drop}%
\end{figure*}

Understanding the deformation of droplets induced by hydrodynamic flows has been a central topic in rheology for almost a century. In 1934, G.\,I. Taylor faced this problem by investigating the deformation of a pure droplet using the approximation of small deformations~\cite{taylor1932viscosity,taylor1934formation}: in his study, Einstein's work for suspensions of solid spheres~\cite{einstein1906new} was extended to the case of droplets with radius $R$ and viscosity $\lambda\mu$ deformed by the action of an external flow with shear intensity $\gammadot$ and viscosity $\mu$ (see Fig.~\ref{fig:drop}).
Droplet deforms due to the competition between viscous forces and surface tension forces resulting from a surface tension $\sigma$: the dimensionless Capillary number is then introduced as $\Ca=\mu R\gammadot /\sigma$ and the droplet deformation is quantified by the Taylor index (see Fig.\ref{fig:drop}).
Taylor's work, whose results are valid in the perturbative limit of $\Ca\ll 1$, proved to be a fundamental approach to the problem, paving the way for several other studies which refine and extend the investigation in many directions, e.g. by considering higher orders in the perturbation theory~\cite{chaffey1967second,barthes1973deformation}, analyzing the time-dependent deformation properties~\cite{cox1969deformation,frankel1970constitutive}, introducing the effects of different types of flows ~\cite{hakimi1980effects,youngren1976shape}, plugging additional complexities in the suspension system~\cite{greco2002drop,vananroye2006effect}.
Various reviews have been written on the topic, covering experimental, theoretical and numerical aspects~\cite{rallison1984deformation,stone1994dynamics,fischer2007emulsion,cristini2004theory}. \\
Droplet dynamics is a very complex non-linear problem and, in many instances, it is desirable to possess a simpler phenomenological description via reduced models. Maffettone \& Minale (MM)~\cite{maffettone1998equation} proposed an equation for the evolution of a droplet which is assumed to be ellipsoidal at all times. The droplet shape is described via a second order tensor $\vec{S}$ evolving with the equation 
\begin{eqnarray}\label{MMmodeladim}
    \frac{d\vec{S}}{dt'}-\Ca\,\left(\boldsymbol{\Omega}\cdot\boldsymbol{S}-\boldsymbol{S}\cdot\boldsymbol{\Omega}\right)= -\fmmone(\lambda)\left[\boldsymbol{S}-g\left(\boldsymbol{S}\right)\boldsymbol{I}\right]
    + \Ca\,\fmmtwo(\lambda)\left(\boldsymbol{E}\cdot\vec{S}+\vec{S}\cdot\boldsymbol{E}\right) \ ,
\end{eqnarray}
where $\boldsymbol{E}$ and  $\boldsymbol{\Omega}$ represent the symmetric and asymmetric velocity gradient tensors respectively, $\boldsymbol{I}$ the unity tensor and $g\left(\vec{S}\right)$ a non-linear function of the $\vec{S}$ tensor~\cite{maffettone1998equation}. Time $t'={t}/{\tau}$ is made dimensionless via the droplet characteristic time $\tau=\mu R / \sigma$. The model parameter $\fmmone(\lambda)$ can be linked to the loading time of the droplet $\tau_{\mbox{\tiny load}}$, i.e., the time the droplet takes to reach a stationary state, while  $\fmmtwo(\lambda)$ is linked to the deformation of the droplet induced by the flow~\cite{maffettone1998equation}. The dependence of these two model parameters on the viscosity ratio $\lambda$ is fixed by requiring that in the limit of small $\Ca$ the model recovers the known results of perturbation theory~\cite{rallison1980note}, in agreement with Taylor's theory of small deformation~\cite{taylor1932viscosity,taylor1934formation}. 
The MM model gained great success thanks to its ability to reproduce experimental data~\cite{maffettone1998equation,torza1972particle,guido1998three,bentley1986experimental} and it has been extended in many directions, e.g. to account for non-Newtonian effects based on experimental and theoretical inputs~\cite{maffettone2004ellipsoidal,minale2004deformation,greco2002drop,guido2003deformation}, to investigate the deformation of a drop in confined flows~\cite{minale2008phenomenological,minale2010microconfined} or to account for the tank-treading motion of the membrane in a soft suspension~\cite{art:arora04}.
Based on experimental observations of droplet deformation up to subcritical regimes~\cite{grace1982dispersion,bentley1986experimental,almusallam2000constitutive}, some models were also proposed to predict the non-linear response at high values of capillary number~\cite{almusallam2000constitutive,jackson2003model,yu2003ellipsoidal}. 
An extensive review on the variety of reduced models for droplet dynamics can be found in~\cite{minale2010models}.\\ 
When designing a reduced model for droplet dynamics, it is a standard practice to determine the model coefficients by requiring a theoretical matching with the asymptotic perturbative expansion in the capillary number $\Ca$ coming from Stokes equations. In this paper we take a different route, developing an alternative strategy to construct the model parameters via direct matching with numerical simulations of Stokes equations at $\Ca \sim {\cal O}(1)$. Specifically, we will study the process of droplet deformation subject to a simple shear flow and construct an extended MM (EMM) model with parameters $f_1^{(\mbox{\tiny EMM})}(\lambda,\Ca)$ and $f_2^{(\mbox{\tiny EMM})}(\lambda,\Ca)$ capable of reproducing loading times and non-linear rheology at $\Ca \sim {\cal O}(1)$. Thus, we retain the attractiveness and ease of a single second-order tensorial model, plugging a functional dependency on $\Ca$ in the model coefficients. The EMM model is then studied at changing $\lambda$ up to the subcritical regime~\cite{grace1982dispersion,bentley1986experimental,cristini2003drop}. The numerical solutions of Stokes equations are obtained via numerical simulations employing the Immersed Boundary - Lattice Boltzmann (IB-LB) method~\cite{feng2004immersed,zhangImmersedBoundaryLattice2007,dupin2007modeling,book:kruger}, a hybrid numerical method widely employed for simulating the dynamics of deformable soft interfaces~\cite{thesis:kruger,li2019finite,guglietta2020effects,li2021similar,kruger2013crossover,art:kruger14deformability, gekle2016strongly}. Our study is also instrumental to further strengthen the computational applicability of this numerical technique in situations aimed at capturing the features of droplet deformations in subcritical regimes. \\
The paper is organized as follows: in Sec.~\ref{sec:statement} we review the basic ingredients of the MM model and summarize relevant model predictions that will be relevant for our analysis; in Sec.~\ref{sec:methods} we describe the IB-LB method used in this work to simulate the dynamics of a single droplet in simple shear flow; results and discussions are presented in Sec.~\ref{sec:results}; conclusions and a summary of our results will be presented in Sec.~\ref{sec:conclusions}. \\
\section{Maffettone-Minale (MM) Model and proposed extension (EMM) }\label{sec:statement}
The Maffettone-Minale (MM) model~\cite{maffettone1998equation} is structured by assuming that the droplet retains an ellipsoidal shape at all times and the second-order tensor $\vec{S}$ used to describe its shape is assumed to evolve due to two main competing effects: the drag exerted by the inner and outer fluid and the surface tension tendency to restore the spherical geometry; while being deformed, the droplet is assumed to further conserve the volume.
The dimensionless equation describing the time evolution of $\vec{S}$ proposed by MM follows as:
\begin{eqnarray}\label{mmmodeladim}
    \frac{d\vec{S}}{dt'}-\Ca\,\left(\boldsymbol{\Omega}\cdot\boldsymbol{S}-\boldsymbol{S}\cdot\boldsymbol{\Omega}\right)= -f_1 \left[\boldsymbol{S}-g\left(\boldsymbol{S}\right)\boldsymbol{I}\right]+\Ca\,f_2 \left(\boldsymbol{E}\cdot\vec{S}+\vec{S}\cdot\boldsymbol{E}\right) \ ,
\end{eqnarray}
where $\boldsymbol{E}= \frac{1}{2}\left(\boldsymbol\nabla\vec{u}+\boldsymbol\nabla\vec{u}^T\right) $ and $\boldsymbol{\Omega}=\frac{1}{2}\left(\boldsymbol\nabla\vec{u}-\boldsymbol\nabla\vec{u}^T\right)$ are the symmetric and asymmetric parts of the velocity gradient $\boldsymbol\nabla\vec{u}$. The LHS of Eq.~\eqref{mmmodeladim} contains the Jaumann derivative and $g\left(\vec{S}\right)=\frac{3 III_S}{II_S}$ is an additional non-linear function plugged in to satisfy the constraint of volume conservation, with:
\begin{equation}\label{tensorinvariants}
    II_S=\frac{1}{2}\left[\trace{\boldsymbol{S}}^2-\trace{\boldsymbol{S}^2}\right]\ , \qquad III_S=\det(\boldsymbol{S})\ ,
\end{equation}
and $\Ca$ is the capillary number:
\begin{equation}
\Ca=\frac{\mu R \dot{\gamma}}{\sigma}\ ,
\end{equation}
where $\mu$ is the viscosity of the fluid, $R$ is the radius of the droplet at rest and $\sigma$ is the surface tension.\\
In order to evaluate the coefficients $f_1$ and $f_2$, MM expand the shape of the drop $\vec{S}$ in the capillary number $\Ca$ as follows:
\begin{equation}\label{firstordertensor}
\vec{S}=\boldsymbol{I}+\Ca\vec{F}\ ,
\end{equation}
with $\vec{F}$ capturing the deformations ``beyond sphericity" whose magnitude is expressed by the capillary number $\Ca$, and they substitute it in Eq.~\eqref{mmmodeladim}, obtaining: 
\begin{equation}\label{MMeps}
\frac{d\vec{F}}{dt}=-f_1\vec{F}+2f_2\vec{E} \ .
\end{equation}
They finally compare the latter with the expansion to the first order in $\Ca$ (see refs.\cite{maffettone1998equation,rallison1980note} and references therein for further details on the perturbative approaches at the problem), finding: 
\begin{align}\label{f1f2mm}
&f_1=\fmmone(\lambda)=\frac{40\left(\lambda+1\right)}{\left(2\lambda+3\right)\left(19\lambda+16\right)}\ ,\nonumber \\
&f_2=\fmmtwo(\lambda)=\frac{5}{2\lambda+3}\ .
\end{align}
The information on the deformation can be retrieved from the eigenvectors of $\vec{S}$, whose eigenvalues correspond to the square of the ellipsoid semiaxis $L^2, B^2, W^2$. The major ($L$) and minor ($B$) semiaxis in the shear plane are used to compute the Taylor index $D=\frac{L-B}{L+B}$ as well as the inclination angle $\theta$ (see Fig.~\ref{fig:drop}). 
We are interested in the deformation $D$ and the inclination angle $\theta$ in simple shear flow, whose matrices $\vec{E}$ and $\vec{\Omega}$ are given by: 
\begin{equation}\label{sheartensors}
    \boldsymbol{E}=\frac{1}{2}\begin{pmatrix}0 & 1 & 0\\1 & 0 & 0\\0 & 0 & 0\\ \end{pmatrix} \qquad
    \boldsymbol{\Omega}=\frac{1}{2}\begin{pmatrix}0 & 1 & 0\\-1 & 0 &0\\0 & 0 & 0  \end{pmatrix} \ .
\end{equation}
By substituting them in Eq.~\eqref{mmmodeladim} and imposing $\frac{d\vec{S}}{dt}=\vec{0}$, the model provides the following analytical formulae for both $D$ and $\theta$ at the stead-state~\cite{maffettone1998equation}:
\begin{equation}\label{eq:stationaryD}
    D=\frac{\sqrt{f_1^2+\Ca^2}-\sqrt{f_1^2+\Ca^2-f_2^2\Ca^2}}{f_2\Ca} \ ,
\end{equation}
\begin{equation}\label{stationarytheta}
\theta=\frac{1}{2}\arctan\left(\frac{f_1}{\Ca}\right).
\end{equation}
The model parameter $f_1$ physically represents the inverse loading time $\tau_{\mbox{{\tiny load}}}$, i.e., the time the droplet takes to reach the stationary state. As remarked above, however, this coefficient is fixed by requiring a theoretical matching with the perturbative results coming from Stokes equation, hence the validity of Eqs.~\eqref{eq:stationaryD}~\eqref{stationarytheta} remains restricted to the small $\Ca$ regime.  An explicit mathematical formula for $f_{1,2}$ valid for all $\Ca$ would require a non perturbative analytical solution of Stokes equations at finite $\Ca$, which is an unfeasible task. Nonetheless, $f_1^{-1}$ is a measurable time, e.g. it can be measured from a direct analysis of the time evolution of the Taylor index $D$. One could therefore think of modifying $f_1$ by directly measuring the loading time via dedicated numerical simulations; one can then use the steady state results from droplet deformation to modify $f_2$ in such a way that the steady deformation predicted by the MM model in Eq.~\eqref{eq:stationaryD} matches the one that is measured in numerical simulations. In this way, one can define an ``Extended Maffettone-Minale" model (hereafter denoted with EMM) with model coefficients $f_1^{(\mbox{\tiny EMM})}$ and $f_2^{(\mbox{\tiny EMM})}$ that naturally acquire a dependency on $\Ca$
\begin{equation}
f_1=f_1^{(\mbox{\tiny EMM})}(\lambda,\Ca)\ , \hspace{.2in} f_2=f_2^{(\mbox{\tiny EMM})}(\lambda,\Ca) \ ,
\end{equation}
and will reduce to the original MM coefficients when $\Ca \rightarrow 0$
\begin{equation}
\lim_{\Ca \rightarrow 0} f_{1,2}^{(\mbox{\tiny EMM})}(\lambda,\Ca)=f^{(\mbox{\tiny MM})}_{1,2}(\lambda).
\end{equation}
The resulting EMM model will then possess (by construction) more realistic loading times and a more realistic steady deformation at finite $\Ca$. This is the strategy that we will pursue to extend the MM model. 

\section{Numerical Simulations: Immersed Boundary - Lattice Boltzmann (IB-LB) Method}\label{sec:methods}
Numerical simulations are needed for the characterization of the droplet deformation process under shear flow. In general, various numerical methods have been developed to simulate bulk viscous flows in presence of deformable soft suspensions, like boundary element methods~\cite{rallison1978numerical,pozrikidis1992boundary,cristini2003drop,art:gounley16}, volume-of-fluid methods~\cite{li2000numerical} and lattice Boltzmann methods~\cite{book:kruger}. 
In this work we use the  Immersed Boundary-Lattice Boltzmann (IB-LB) method.
We remark that simulations of droplets in viscous flows can also be tackled via LB in conjunction with some non-ideal interface force model, like the Shan-Chen model~\cite{shan1993lattice,shan1994simulation}, the Free-Energy model~\cite{swift1995lattice,swift1996lattice}, the color gradient model~\cite{liu2012three} or the entropic model~\cite{chikatamarla2015entropic}. These models, however, are diffuse interface models and their use would require a precise convergence to the sharp-interface limit of hydrodynamics, whereas IB-LB naturally preserves this limit. Additionally, on a future perspective, we plan to use the IB-LB tool to characterise droplet dynamics in a generic time-dependent turbulent strain matrix. To do this, improved boundary conditions need to be implemented at the boundaries of the computational domain~\cite{milan2020sub}, and the use of a ``basic'' LB (like the one we use in our IB-LB) is more suited than a non-ideal diffuse interface LB method.\\  
The IB-LB method has already been used in previous works for investigating the dynamics of droplets and viscoelastic capsules~\cite{thesis:kruger,li2019finite,guglietta2020effects,li2021similar,kruger2013crossover,art:kruger14deformability,gekle2016strongly}, hence we only recall the relevant features of the method (the interested reader can found more details in the aforementioned works and references therein). The IB-LB method hinges on the Immersed Boundary (IB) model that couples the interface of the droplet to the fluid, and on the Lattice Boltzmann method (LB) that simulates the bulk viscous flows. The LB method is a kinetic approach to simulate hydrodynamics~\cite{book:kruger,succi2018lattice}: we will briefly illustrate its working principles taking into consideration the bulk fluid in the outer droplet region, with macroscopic density $\rho$, velocity $\vec{u}$ and viscosity $\mu$ (the same reasoning applies to the bulk fluid inside the droplet). The target equations for the LB method are the continuity equation and the Navier-Stokes equations 
\begin{equation}\label{eq:continuity}
\frac{\partial \rho}{\partial t} + \boldsymbol{\nabla}\cdot (\rho \vec{u})=0\ ,
\end{equation}
\begin{equation}\label{eq:NS}
\rho \left(\frac{\partial\vec{u}}{\partial t}+(\vec{u}\cdot\pmb{\nabla})\vec{u}\right)=-\pmb{\nabla} p+\mu \pmb{\nabla}^2\vec{u}+\vec{F}_{\tiny{\mbox{ext}}}\ ,
\end{equation}
where $\vec{F}_{\tiny{\mbox{ext}}}$ accounts for external forces. Instead of solving the hydrodynamic equations directly, the LB method evolves in time the probability distribution functions that can stream along a finite set of directions, according to the finite set of kinetic velocities $\vec{c}_i$ ($i=0...Q-1$). We implement the so-called D3Q19 velocity scheme, featuring $Q=19$ directions~\cite{book:kruger,succi2018lattice}, with kinetic velocities 
\begin{widetext}
\begin{equation}
\vec{c}_i=
\begin{cases} 
(0,0,0) & i=0 \\ 
(\pm \Delta x/\Delta t,0,0), (0,\pm \Delta x/\Delta t,0), (0,0,\pm \Delta x/\Delta t) & i=1-6 \\ (\pm \Delta x/\Delta t,\pm \Delta x/\Delta t,0), (0,\pm \Delta x/\Delta t,\pm \Delta x/\Delta t), (\pm \Delta x/\Delta t,0,\pm \Delta x/\Delta t) & i=7-18 \ , \end{cases}  
\end{equation}
\end{widetext}
where $\Delta x$ is the lattice spacing and $\Delta t$ is the discretised time step.
The probability distribution function on the fluid (Eulerian) node with coordinates $\vec{x}$ moving with discrete velocity $\vec{c}_i$ at time $t$ is represented by $\effi(\vec{x},t)$, and the LB equation reads:
\begin{equation}\label{LBMEQ}
\effi(\vec{x}+\vec{c}_i\Delta t, t+ \Delta t) - \effi(\vec{x}, t) =\Delta t \left[\Omega_i(\vec{x},t) + S_i(\vec{x},t)\right]\ ,
\end{equation}
where the left-hand side represents the streaming along the direction $i$, while in the right-hand side we find the collision term. 
In this work, we implemented the standard Bhatnagar-Gross-Krook (BGK) collision operator~\cite{qian1992lattice,book:kruger}:
\begin{equation}\label{eq:collision}
\Omega_i(\vec{x},t)=-\frac{1}{\tau_{\mbox{\tiny LB}}}[\effi(\vec{x}, t) - \effi^{(\mbox{\tiny eq})}(\vec{x}, t)]\ ,
\end{equation}
expressing the relaxation towards a local equilibrium $\effi^{(\mbox{\tiny eq})}(\vec{x}, t)$ with characteristic relaxation time $\tau_{\mbox{\tiny LB}}$. 
The local equilibrium depends on $\vec{x}$ and $t$ via the density $\rho(\vec{x},t)$ and velocity field $\vec{u}(\vec{x},t)$~\cite{qian1992lattice}:
\begin{equation}
\effi^{(\mbox{\tiny eq})}(\vec{x},t)= w_i\rho\left(1+\frac{\vec{u}\cdot\vec{c}_i}{c_s^2}+\frac{(\vec{u}\cdot\vec{c}_i)^2}{2c_s^4}-\frac{\vec{u}\cdot\vec{u}}{c_s^2}\right)\ ,
\end{equation}
where $w_i=w(|\vec{c}_i|^2)$ are suitable weights, such that $w_0=1/3$, $w_{1-6}=1/18$, $w_{7-18}=1/36$ and $\cs=\Delta x/\Delta t\sqrt{3}$ is the speed of sound.
It is worth noting that more sophisticated collision operators might be implemented (see ref.~\cite{book:kruger}): however, since we only need to retrieve the incompressible hydrodynamics with one free parameter (i.e., the viscosity of the fluid), the BGK collision operator is enough to fulfill these requirements.

The relaxation process is supplemented by the presence of external forces implemented in the LB via the source term $S_i(\vec{x},t)$ according to the so-called Guo scheme~\cite{guo2002discrete}:
\begin{equation}
S_i(\vec{x},t)=\left(1-\frac{\Delta t}{2\tau_{\mbox{\tiny LB}}}\right)\frac{w_i}{c_s^2}\left[\left(\frac{\vec{c}_i\cdot \vec{u}}{c_s^2}+1\right)\vec{c}_i-\vec{u}\right]\cdot\vec{F}_{\tiny{\mbox{ext}}}\ .
\end{equation}
 The populations $\effi$ are used to compute the hydrodynamic density and momentum fields:
\begin{subequations}\label{eq:density_velocity_LB}
\begin{align}\label{eq:density_LB}
&\rho(\vec{x}, t) = \sum_{i} \effi(\vec{x}, t)\; , \\
&\rho\vec{u}(\vec{x}, t) = \sum_{i} \vec{c}_i \effi(\vec{x}, t)+\frac{\vec{F}_{\tiny{\mbox{ext}}}\Delta t}{2}\ , \label{eq:velocity_LB}
\end{align}
\end{subequations}
where the velocity contains the so-called half-force correction, coming from the choice of the Guo forcing~\cite{book:kruger,guo2002discrete}.
Details on the computation of the force $\vec{F}_{\tiny{\mbox{ext}}}$ are given below. The LB equation recovers the continuity and Navier-Stokes equations (see Eq.~\eqref{eq:continuity} and Eq.~\eqref{eq:NS}, respectively) when fluctuations around the local equilibrium are small. In such a case, hydrodynamic description is granted with viscosity $\mu= \rho \cs^2\left(\tau_{\mbox{\tiny LB}}-\frac{\Delta t}{2}\right)$ and pressure $p=c_s^2 \rho$. 

The interface of the droplet is represented by a finite set of Lagrangian points linked together to build a 3D triangular mesh: on each triangular element, the stress is given by the surface tension $\vec{\tau}_\sigma^{(\vec{S})}=\sigma\boldsymbol{I}$. 
Details on the computation of the force $\vec{\varphi}_i$ on the $i-$th Lagrangian point from the stress tensor are given in~\cite{guglietta2020effects}. The force $\vec{\varphi}_i$ is thus communicated from the Lagrangian point with coordinates $\vec{r}_i$ to the surrounding Eulerian (fluid) nodes, while the velocity of the Lagrangian point $\dot{\vec{r}}_i(t)$ is retrieved from the fluid velocity computed on the neighbouring Eulerian nodes (see Eq.~\eqref{eq:velocity_LB}). Such a two-way coupling is handled via an interpolation according to the IB method~\cite{peskin2002immersed,book:kruger}:
\begin{subequations}
\begin{align}\label{eq:ibm_force_d}
&\vec{F}_{\tiny{\mbox{ext}}}(\vec{x},t) =  \sum_i\vec{\varphi}_i(t)\Delta(\vec{r}_i-\vec{x})\ , \\
&\dot{\vec{r}}_i(t) = \sum_{\vec{x}} \vec{u}(\vec{x},t)\Delta(\vec{r}_i-\vec{x})\Delta x^3\ ,
\end{align}    
\end{subequations}
where the $\Delta$ function is the discretised Dirac delta, which can be factorised in the three interpolation stencils $\Delta(\vec{x}) = \phi(x)\phi(y)\phi(z)/\Delta x^3$~\cite{book:kruger}. In this work, we implemented the 4-points interpolation stencil~\cite{book:kruger}:\begin{equation}
\phi(x) = \begin{cases}
\frac{1}{8}\left(3-2\vert x\vert + \sqrt{1+4\vert x\vert-4x^2} \right) & 0\le \vert x\vert\; ;\\
\frac{1}{8}\left(5-2\vert x\vert - \sqrt{-7+12\vert x\vert-4x^2} \right) & \Delta x\le \vert x\vert\le 2\Delta x\; ; \\
0 & 2\Delta x\le \vert x\vert\; .
\end{cases}
\end{equation}

IB-LB numerical simulations are performed in a 3D box with size $\Lx\times\Ly\times\Lz=120\times 200\times 120$ lattice units. The linear shear flow is generated via the bounce-back method for moving walls~\cite{book:kruger} that are located in $y=\pm\Ly/2$ with velocity $\Uw=(\pm \dot{\gamma}\Ly/2,0,0)$, respectively, while periodic boundary conditions are set along the $x$ and $z$ directions. The channel width $\Ly$ is chosen to be large enough to avoid effects of confinement on the droplet deformation~\cite{guido2011shear}. 
We are interested in studying the deformation of the droplet in a fully developed flow rather than considering the transitional and startup effects of an ambient shear flow accelerated from zero to the final velocity;
considering this, the droplet is placed in the centre of the 3D box only once the linear shear flow is fully developed.
This is a standard initialization setup for both experimental and numerical shear flow studies on droplets~\cite{renardy2008effect}.\\
The interface of the droplet is discretised with a 3D triangular mesh made of 81920 faces and with radius $R=19$ lattice units.
Such refined mesh with a large number of surface elements was chosen as to not introduce numerical effects for large droplet deformation via convergence studies on both Taylor index and inclination angle.
We further recover Stokes equation and the incompressibility equations by choosing a very small Reynolds number $\Rey=\rho R u/ \mu$ and allowing for small density fluctuations. Simulations are performed in the limit of small Reynolds number (i.e., $\Rey< 10^{-3}$). The information on the major and minor axes is retrieved from the eigenvalues $\mathcal{I}_i$ of the inertia tensor $\vec{\mathcal{I}}$, that is, $L=2\sqrt{5(\mathcal{I}_2+\mathcal{I}_3-\mathcal{I}_1)/(2M)}$ and $B=2\sqrt{5(\mathcal{I}_2+\mathcal{I}_3-\mathcal{I}_1)/(2M)}$, where M is the mass of the drop~\cite{thesis:kruger}. To compute the inclination angle $\theta$, we use again the coordinates of the major eigenvector.

We ran numerical simulations on Graphic Processing Units (GPUs). In particular, we used Nvidia Ampere ``A100'' cards, which allowed us to perform fast simulations: a typical run of $10^7$ time steps (in lattice units) takes about 6 hours.

\begin{figure*}[t!]
\centering
\includegraphics[width=1.0\linewidth]{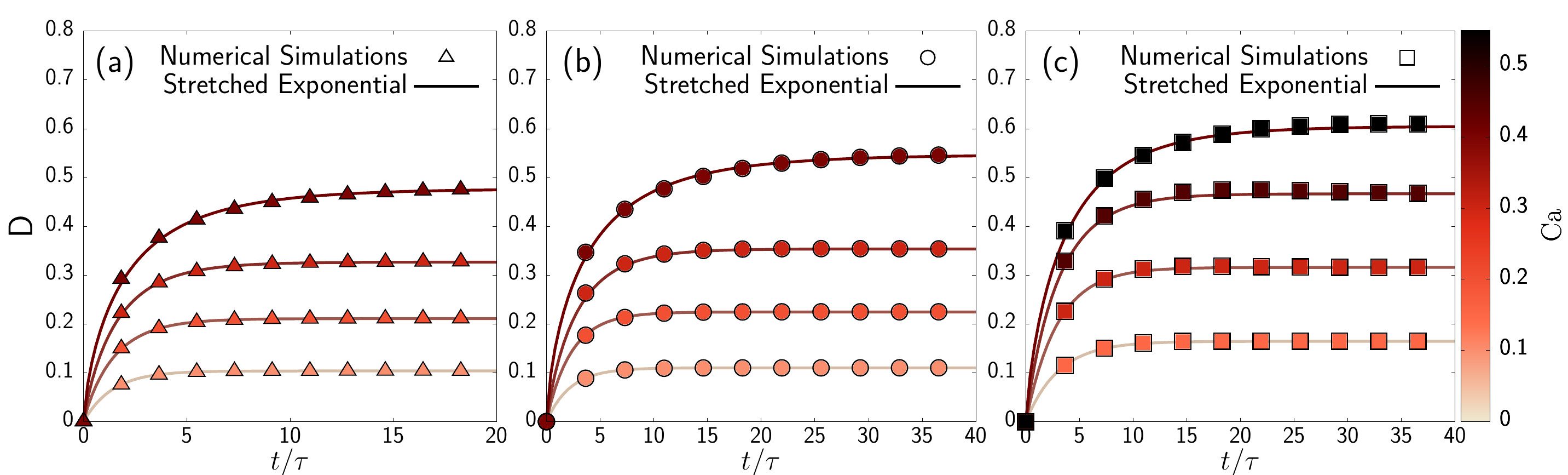}%
\caption{Time evolution of the Taylor index $D$ at increasing $\Ca$ (lighter to darker colors). Time is made dimensionless via droplet time $\tau=\mu R/\sigma$. Numerical simulations data (points) are fitted with a stretched exponential (solid lines, Eq.~\eqref{stretchedexpo}). Different viscosity ratios $\lambda$ are considered. Panel (a): $\lambda=0.2$ and $\Ca=0.1$ (\protect \trianglea), $\Ca=0.2$ (\protect \triangleb),$\Ca=0.3$ (\protect \trianglec), $\Ca=0.4$ (\protect \triangled);
Panel (b): $\lambda=1$ and $\Ca=0.1$ (\protect \circlea), $\Ca=0.2$ (\protect \circleb), $\Ca=0.3$ (\protect \circlec), $\Ca=0.4$ (\protect \circled);
Panel (c): $\lambda=2$ and $\Ca=0.15$ (\protect \squarea), $\Ca=0.3$ (\protect \squareb), $\Ca=0.45$ (\protect \squarec), $\Ca=0.55$ (\protect \squared).}\label{fig:threepanels_transient}
\end{figure*}
\begin{figure*}[t!]
\centering
\includegraphics[width=1.0\linewidth]{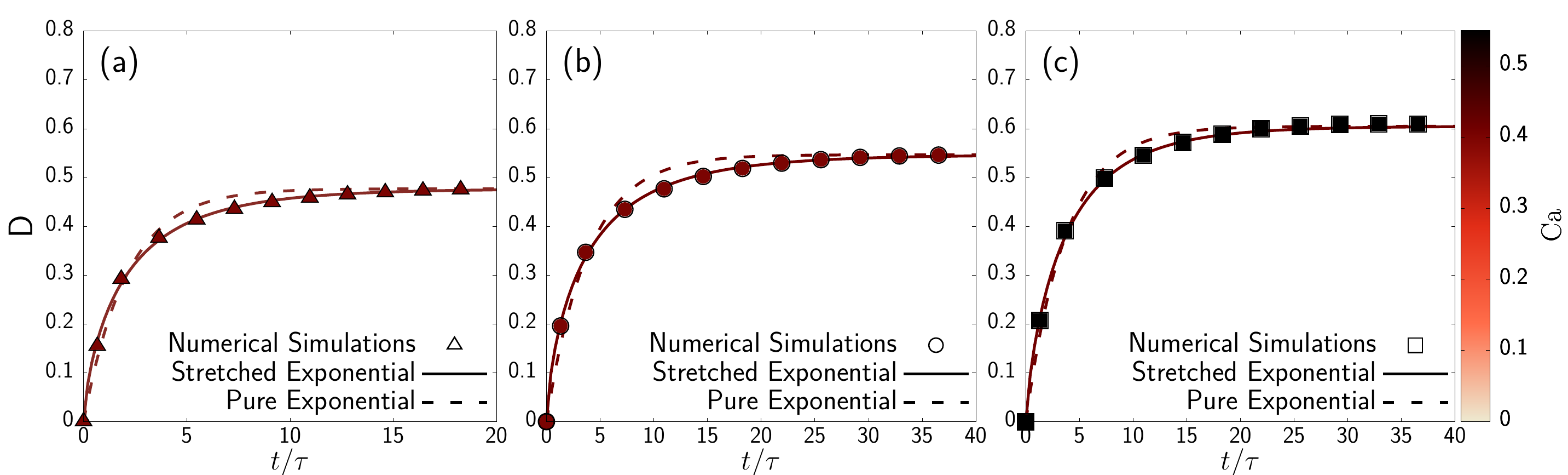}%
\caption{Comparison between numerical simulations data (points), a stretched exponential fit (solid lines, Eq.~\eqref{stretchedexpo} with $\delta \neq 1$) and pure exponential fit (dashed lines, Eq.~\eqref{stretchedexpo} with $\delta=1$) for the time evolution of the Taylor index $D$ for different $\lambda$ and selected values of $\Ca$.
Time is made dimensionless via droplet time $\tau=\mu R/\sigma$.
Panel (a): $\lambda=0.2$, $\Ca=0.4$ (\protect \triangled), $\delta=0.742$;
Panel (b): $\lambda=1$, $\Ca=0.4$ (\protect \circled), $\delta=0.723$;
Panel (c): $\lambda=2$, $\Ca=0.55$ (\protect \squared), $\delta=0.796$.}\label{fig:stretchedcomparison}
\end{figure*}
\begin{figure*}[t!]
\centering
\includegraphics[width=1.0\linewidth]{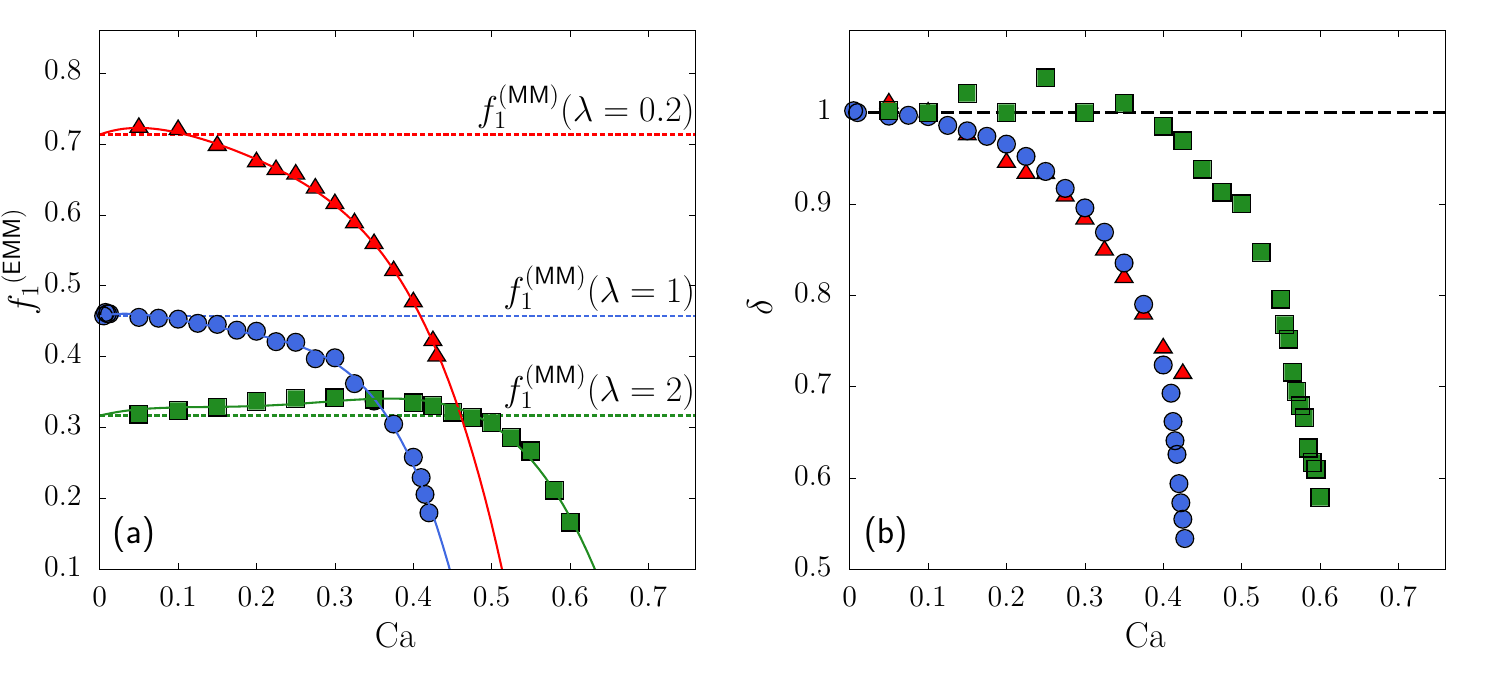}%
\caption{Panel (a): characteristic dimensionless loading frequency $f_1^{(\mbox{\tiny EMM})}$ as a function of the capillary number $\Ca$ (see Eq.~\eqref{stretchedexpo}) for $\lambda=0.2$ (\protect \triangleRED), $\lambda=1$ (\protect \circleBLUE), $\lambda=2$ (\protect \squareGREEN). The horizontal dashed lines indicate the values of $\fmmone(\lambda)$ (see Eq.~\eqref{f1f2mm}), while solid lines refer to a polynomial fit for $f_1^{(\mbox{\tiny EMM})}$  (see Eq.~\eqref{eq:f12EMM} and Table~\ref{table:1}).
Panel (b): stretching factor $\delta$ (see Eq.~\eqref{stretchedexpo}) as a function of $\Ca$ for different values of $\lambda$. The horizontal dashed line is drawn in correspondence of $\delta=1$.}\label{fig:twopanels_deltas_f1}
\end{figure*}
\section{Results and discussions}\label{sec:results}
\begin{figure}[t!]
\centering
\includegraphics[width=0.7\linewidth]{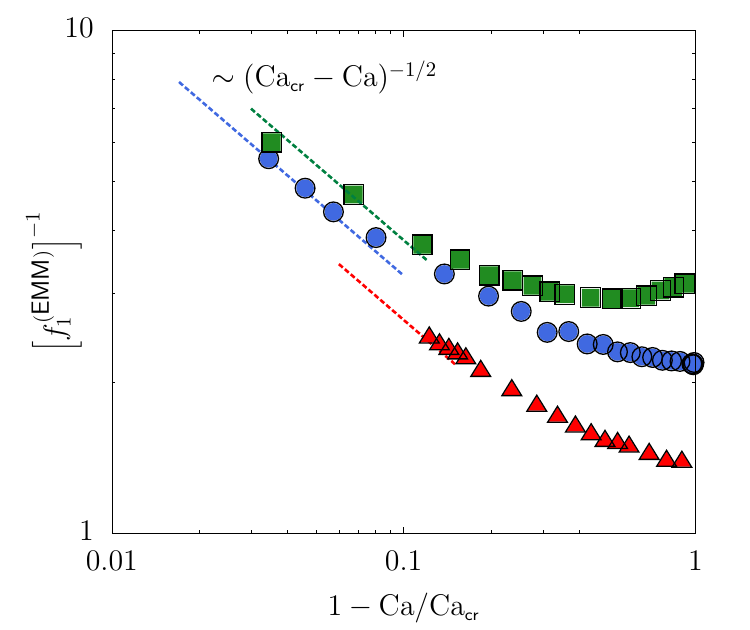}%
\caption{Subcritical scaling for the dimensionless characteristic loading time $\tau_{\mbox{{\tiny load}}}=[f_1^{(\mbox{\tiny EMM})}]^{-1}$ for $\lambda=0.2$ (\protect \triangleRED), $\lambda=1$ (\protect \circleBLUE) and $\lambda=2$ (\protect \squareGREEN). Dashed lines indicate the scaling prediction with exponent $-1/2$ ~\cite{blawzdziewicz2002critical}.}\label{fig:sub-crit_Lambda1}%
\end{figure}

We start our investigation by analyzing the process of droplet deformation for different viscosity ratios $\lambda$. We choose $\lambda$ to span one decade, $\lambda \in [0.2,2]$, which is a realistic range~\cite{lessard2000significance,carvalho2017viscosity} where we can characterize the droplet deformation process from the small deformations up to the breakup occurring at moderate values of capillary number $\Ca$~\cite{barthes1973deformation,grace1982dispersion,cristini2003drop}.\\
In Fig.~\ref{fig:threepanels_transient}, we report the evolution of the Taylor index (see Fig.~\ref{fig:drop}) for different values of the capillary number $\Ca$. Time is made dimensionless with respect to the droplet time $\tau=\mu R / \sigma$. The linearization of the MM model would imply the deformation to exponentially increase with a single characteristic time equal to $f_1^{-1}$ up to the steady deformation value $D_{\infty}$, i.e., $D(t)=D_{\infty}(1-e^{-f_1 t/\tau})$: this behaviour is in fact in good agreement with the numerical data only for small deformations, while for larger deformations we find that the following stretched exponential function
\begin{equation}\label{stretchedexpo}
D(t)=D_{\infty}\left[1-e^{-(f_1 t/\tau)^{\delta}}\right]\ ,
\end{equation}
better adapts to the data, where $\delta$ is the stretching factor used as a fitting parameter. The improved agreement brought by the stretched exponential function is highlighted in Fig.~\ref{fig:stretchedcomparison}. The stretched exponential function is mostly used to describe non-linear relaxation processes~\cite{phillips1996stretched,fedosov2010multiscale} and its basis can be understood in the context of the relaxation of homogeneous glasses and glass-forming liquids~\cite{palmer1984models,potuzak2011topological}, i.e., physical systems that are known to exhibit multiple relaxation times~\cite{jurlewicz1993relationship,elton2018stretched}. Introducing $\delta$ makes it possible to discern a global characteristic time in a system whose evolution process is subjected to multiple time scales~\cite{johnston2006stretched}: for $\delta=1$, we reduce to a linear relaxation process with a single characteristic time, while $\delta<1$ signals the emergence of a spectrum of characteristic times wherein the value $\tau_{\mbox{{\tiny load}}}=\left[f_1^{(\mbox{\tiny EMM})}\right]^{-1}$ is a representative one. The values of $\delta$ and $f_1^{(\mbox{\tiny EMM})}$ that we extracted from this fitting procedure are displayed in Fig.~\ref{fig:twopanels_deltas_f1}, showing that an increase in $\Ca$ results in a decrease of both $\delta$ and $f_1^{(\mbox{\tiny EMM})}$. As the droplet approaches large deformations, it gets closer to the critical point of breakup and the characteristic time for the deformation process  $\tau_{\mbox{{\tiny load}}}$ is expected to get larger, i.e., the characteristic loading frequency $f_1^{(\mbox{\tiny EMM})}$ gets smaller~\cite{blawzdziewicz2002critical,cristini2003drop}. To characterize the approach towards criticality in a more quantitative way, we analyse the subcritical scaling of the loading time $\tau_{\mbox{{\tiny load}}}$, expected to follow a power law $\tau_{\mbox{{\tiny load}}} \sim (\Cacr-\Ca)^{-1/2}$ as showed in~\cite{blawzdziewicz2002critical}. In Fig.~\ref{fig:sub-crit_Lambda1}, we show that the loading time $\tau_{\mbox{{\tiny load}}}$ follows the above mentioned scaling in the subcritical regime, which has been used to evaluate the critical capillary number for breakup $\Cacr$.
In this way, probing subcritical droplet configurations allows one to guess and indirectly measure $\Cacr$, bypassing the limitations of a front-tracking method for the visualization of the break-up process.
We evaluate $\Cacr\sim 0.49$ for $\lambda=0.2$, $\Cacr\sim 0.435$ for $\lambda=1$ and $\Cacr\sim 0.62$ for $\lambda=2$: these values are in good agreement with other evaluations of $\Cacr$ reported in the literature for a droplet in simple shear flow~\cite{grace1982dispersion,cristini2003drop}.\\
\begin{figure*}[t!]
\centering
\includegraphics[width=1.0\linewidth]{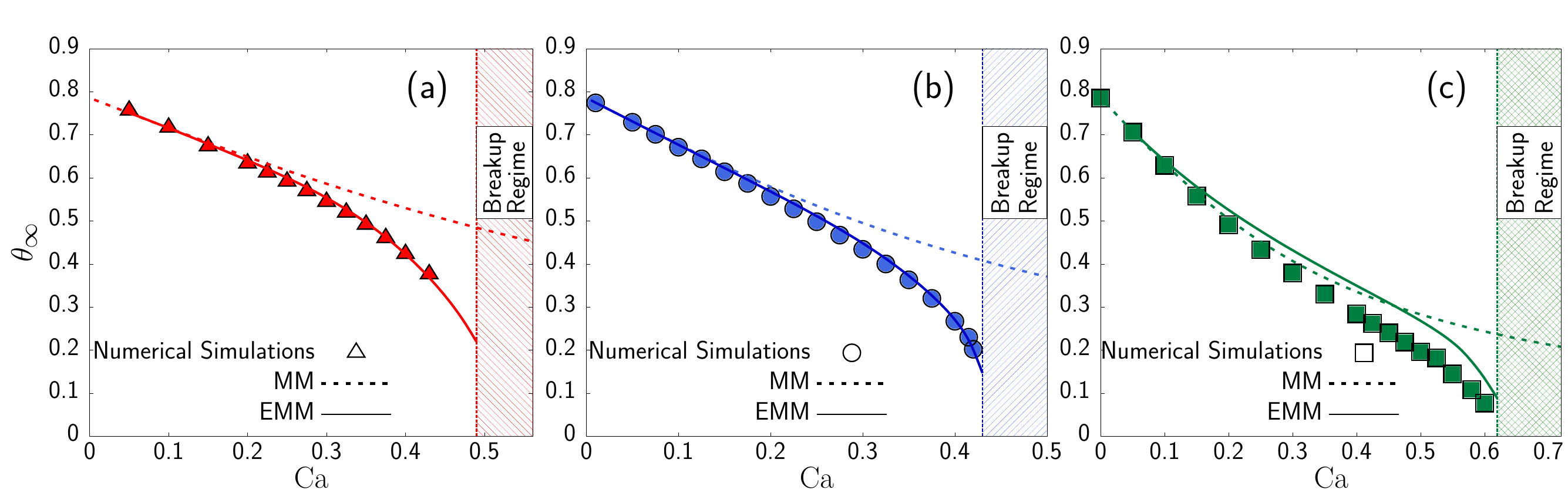}%
\caption{Stationary inclination angles for different $\lambda$. We report results of numerical simulations (points) for   $\lambda=0.2$ (\protect \triangleRED, Panel (a)), $\lambda=1$ (\protect \circleBLUE, Panel (b)), $\lambda=2$ (\protect \squareGREEN, Panel (c)), as well as the prediction of Eq.~\eqref{stationarytheta} with $f_{1}=f_1^{(\mbox{\tiny MM})}$ (MM, dashed lines) and with $f_{1}=f_1^{(\mbox{\tiny EMM})}$ (EMM, solid lines). The colored regions refer to the evaluated breakup regime. }\label{fig:stationary_twopanels_angles}%
\end{figure*}
\begin{figure*}[t!]
\centering
\includegraphics[width=1.0\linewidth]{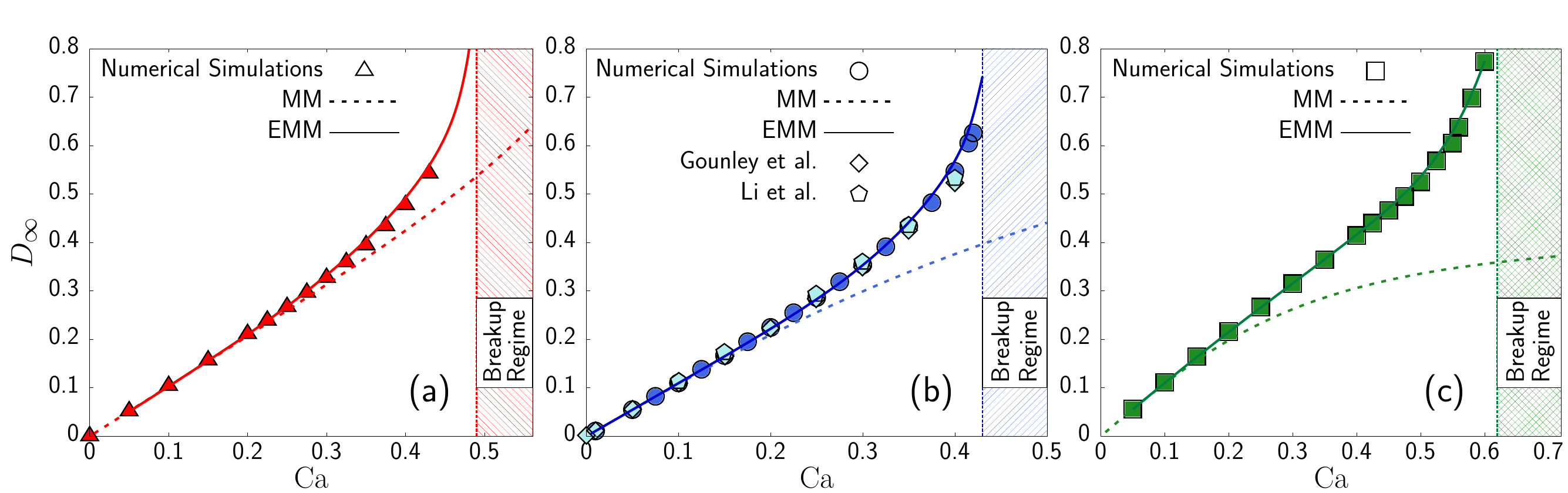}%
\caption{Stationary droplet deformations for different $\lambda$. We report results of numerical simulations (points) for   $\lambda=0.2$ (\protect \triangleRED, Panel (a)), $\lambda=1$ (\protect \circleBLUE, Panel (b)), $\lambda=2$ (\protect \squareGREEN, Panel (c)), as well as the prediction of Eq.~\eqref{eq:stationaryD} with $f_{1,2}=f_{1,2}^{(\mbox{\tiny MM})}$ (MM, dashed lines) and the one with the values $f_{1}=f_1^{(\mbox{\tiny EMM})}$ and the optimal $f_2=f_2^{(\mbox{\tiny EMM})}$ that allow to match the numerical results (EMM, solid lines). In Panel (b) data from Ref.~\cite{art:gounley16} (Gounley et al., \protect \rhombusBLUE, Boundary Element Method) and Ref.~\cite{li2000numerical} (Li et al., \protect \pentagonBLUE, Volume of Fluid) are considered. The colored regions refer to the evaluated breakup regime. \label{fig:stationary_twopanels_defos}}
\end{figure*}
We now consider the prediction of the MM model for the steady inclination angle $\theta_{\infty}$ in Eq.~\eqref{stationarytheta}. In Fig.~\ref{fig:stationary_twopanels_angles}, we report a comparison between the steady values of the inclination angle $\theta_{\infty}$ extracted from numerical simulations (filled points) and those predicted via Eq.~\eqref{stationarytheta} with the coefficient $f_1^{(\mbox{\tiny EMM})}$ being estimated from the loading times (solid lines); the prediction based on Eq.~\eqref{stationarytheta} with $\fmmone$ is also reported (dashed lines). The latter well adapts to the results from numerical simulations for small values of $\Ca$, while a mismatch appears for larger values of $\Ca$. 
The prediction based on the EMM model coefficient $f_1^{(\mbox{\tiny EMM})}$, instead, shows a good match with the numerical data, with only a slight mismatch that is observed for the largest value of viscosity ratio $\lambda$ considered. 
This is due to the development of an overshooting behaviour in the transient dynamics at high $\Ca$ that is not caught via a monotonic function such as Eq.~\eqref{stretchedexpo}. As shown in Fig.~\ref{fig:transient_rk}, the overshoot grows in $\lambda$, driving the slight mismatch observed in Fig.~\ref{fig:stationary_twopanels_angles}.
We have verified that by promoting the fitting function given in Eq.~\eqref{stretchedexpo} to a non monotonous function would improve the accuracy of the loading time evaluation. Overall, the agreement that we observe between the results of numerical simulations and steady state predictions from the EMM model is remarkable. We believe this result is highly non trivial, in that we are able to predict the steady state inclination angle via an independent measurement of the loading time. 

\begin{figure}[t!]
\centering
\includegraphics[width=0.7\linewidth]{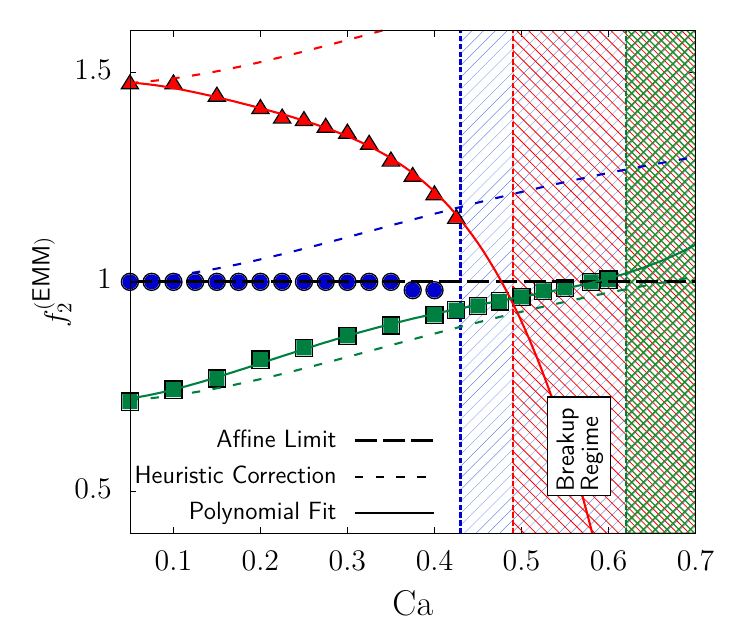}%
\caption{We report the coefficient $f_2^{(\mbox{\tiny EMM})}$ as a function of the capillary number $\Ca$ that allows to match the prediction of the steady deformation in Eq.~\eqref{eq:stationaryD} with numerical simulations data (points) (see Fig.~\ref{fig:stationary_twopanels_defos}) for $\lambda=0.2$ (\protect \triangleRED), $\lambda=1$ (\protect \circleBLUE), $\lambda=2$ (\protect \squareGREEN). The colored regions refer to a range of $\Ca$ where breakup occurs. The heuristic correction suggested and reported in Eq. (34) of Ref.~\cite{maffettone1998equation} is also displayed (Heuristic Correction, colored dashed lines). The limit where the affine deformation takes place ($f_2=1$) is indicated with a black dashed line; solid lines refer to a polynomial fit for $f_2^{(\mbox{\tiny EMM})}$ as in Table~\ref{table:1}.}\label{fig:f2}%
\end{figure}

\begin{figure*}[t!]
\centering
\includegraphics[width=1.0\linewidth]{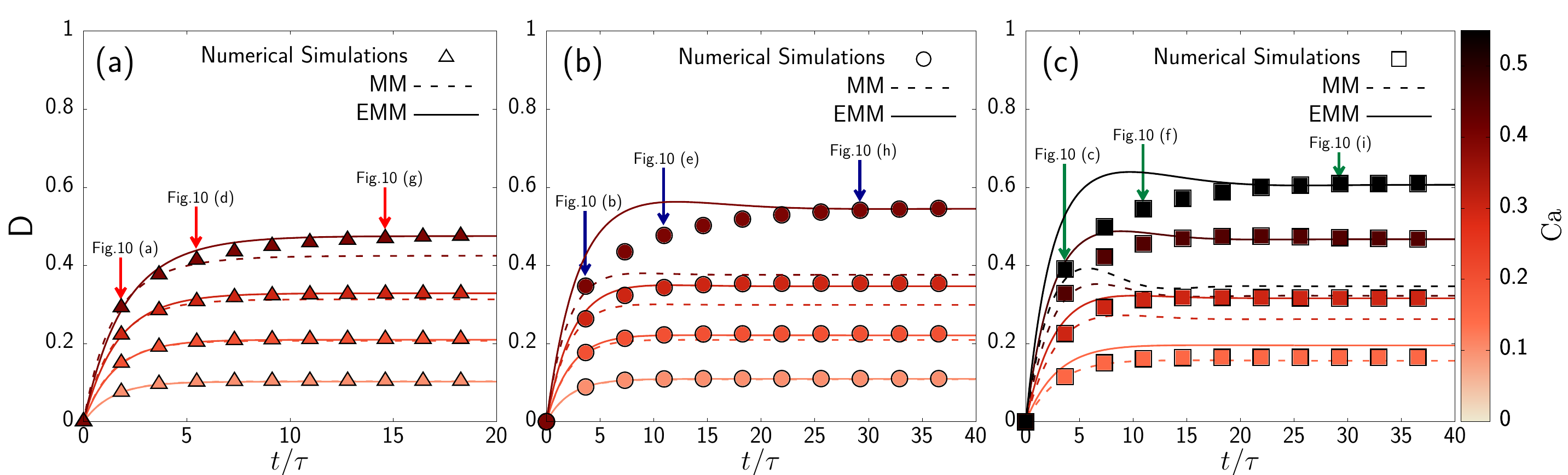}%
\caption{Time evolution of the Taylor index for different $\lambda$ and  $\Ca$. Numerical simulations data (points) are compared with the original MM model as in Eq.~\eqref{mmmodeladim} with $f_{1,2}=f_{1,2}^{(\mbox{\tiny MM})}$ (MM, dashed lines). We also report the predictions of Eq.~\eqref{mmmodeladim} with $f_{1,2}=f_{1,2}^{(\mbox{\tiny EMM})}$ (EMM, solid lines) that we evaluated from the loading times and the steady deformations (see Figs.~\ref{fig:twopanels_deltas_f1} and \ref{fig:f2}).
Panel (a): $\lambda=0.2$ and $\Ca=0.1$ (\protect \trianglea), $\Ca=0.2$ (\protect \triangleb),$\Ca=0.3$ (\protect \trianglec), $\Ca=0.4$ (\protect \triangled);
Panel (b): $\lambda=1$ and $\Ca=0.1$ (\protect \circlea), $\Ca=0.2$ (\protect \circleb), $\Ca=0.3$ (\protect \circlec), $\Ca=0.4$ (\protect \circled);
Panel (c): $\lambda=2$ and $\Ca=0.15$ (\protect \squarea), $\Ca=0.3$ (\protect \squareb), $\Ca=0.45$ (\protect \squarec), $\Ca=0.55$ (\protect \squared).}\label{fig:transient_rk}%
\end{figure*}

We now want to consider the prediction for the steady deformation in Eq.~\eqref{eq:stationaryD} and show the strategy adopted to compute $f_2$. Once we know the value of $f_1^{(\mbox{\tiny EMM})}$, we can look for an optimal value of $f_2=f_2^{(\mbox{\tiny EMM})}$ such that the prediction of the deformation in Eq.~\eqref{eq:stationaryD} matches the data coming from numerical simulations for finite $\Ca$. This matching procedure is illustrated in Fig.~\ref{fig:stationary_twopanels_defos}, where we report the results for the steady deformation extracted from the numerical simulations and the prediction based on Eq.~\eqref{eq:stationaryD} using the MM model parameters $\fmmonetwo$ (dashed lines). 
To strengthen our numerical analysis, comparisons with results based on different numerical methods~\cite{art:gounley16,li2000numerical} are also reported for the equiviscous case as shown in Panel (b).
As already observed for the steady inclination angle (see Fig.~\ref{fig:stationary_twopanels_angles}), the MM prediction in Eq.~\eqref{eq:stationaryD} with $f_{1,2}=\fmmonetwo$ well adapts to the data for small $\Ca$, while a mismatch emerges at $\Ca \sim {\cal O}(1)$. Our working strategy is then to identify, for each value of $\Ca$, the value of $f_2^{(\mbox{\tiny EMM})}$ that allows Eq.~\eqref{eq:stationaryD} to perfectly match with the simulation data.
We find that this matching is possible for the values of $\lambda$ that we considered (solid lines). The values of $f_2^{(\mbox{\tiny EMM})}$ that we extracted with this procedure are reported in Fig.~\ref{fig:f2} and also compared with an heuristic $\Ca-$dependent formula proposed by MM (Eq.(34) in~\cite{maffettone1998equation}).
The latter compares very well with our proposed extensions only for the case $\lambda>1$, where we report a very similar increase in $f_2$ at increasing $\Ca$. Qualitatively, the model parameter $f_2$ is expected to acquire a dependence on $\Ca$, in that higher values of $\Ca$ would imply a droplet deformation process mainly driven by the viscous forces, while surface tension effects become negligible in such situation. More quantitatively,  the model parameter $f_2$ can be viewed as an ``affinity" parameter~\cite{windhab2005emulsion} accounting for an additional surface slip which effectively removes part of the droplet elongation, whereas the full rigid body rotation is retained coherently with the rotating flow~\cite{gordon1972anisotropic}.  When $\lambda\ne 1$, a common behaviour is shown in the approaching of the affine limit $f_2=1$ at higher $\Ca$: the affine motion of the interface changes with respect to the magnitude of the capillary number and the value $f_2=1$ is approached near or at the corresponding critical configuration for the droplet, signaling that an affine motion is ultimately required for the breakup~\cite{elemans1993transient,meijer1994mixing}. We also notice that for the case $\lambda=1$ we have $\fmmtwo=1$, so that in the case of equiviscous fluids the effective slip at the interface is lacking and the droplet is already in the affine configuration at small $\Ca$. We have fitted the coefficients $f_{1,2}^{(\mbox{\tiny EMM})}$ with a polynomial expansion in the capillary number $\Ca$:
\begin{equation}\label{eq:f12EMM}
    f_{1,2}^{(\mbox{\tiny EMM})}(\lambda,\Ca)=\sum_{i} a_i^{(1,2)}(\lambda)\Ca^i\ ,
\end{equation}
in Tab.~\ref{table:1} we report details for a forth-order fit for $\lambda=0.2,1,2$.
\begin{table}[t]
\centering
\setlength{\tabcolsep}{4pt}
\begin{tabular}{c|c c|c c|c c}
\multicolumn{1}{c|}{\multirow{2}{*}{}}&\multicolumn{2}{c|}{$\lambda=0.2$}&\multicolumn{2}{c|}{$\lambda=1$}&\multicolumn{2}{c}{$\lambda=2$} \\
$i$ & $a_i^{(1)}$ & $a_i^{(2)}$ & $a_i^{(1)}$ & $a_i^{(2)}$ & $a_i^{(1)}$ & $a_i^{(2)}$\\ [0.5ex] 
 \hline
 0 & 0.713 & 1.470 & 0.457 & 1 & 0.317 & 0.714 \\ 
 1 & 0.416 & 0.347 & 0.235 & 0 & 0.270 & -0.028 \\
 2 & -5.333 & -6.040 & -4.546 & 0 & -2.283 & 3.996 \\
 3 & 16.776 & 20.021 & 20.536 & 0 & 8.338 & -9.188 \\
 4 & -24.339 & -27.703 & -34.798 & 0 & -9.91 & 6.611 \\
\end{tabular}
\caption{Values of the interpolated coefficients $a_i^{(1,2)}(\lambda)$  for $f_{1,2}^{(\mbox{\tiny EMM})}$ (Eq.~\eqref{eq:f12EMM}, with forth order interpolation).}
\label{table:1}
\end{table}

\begin{figure*}[t!]
\centering
\includegraphics[width=0.755\linewidth]{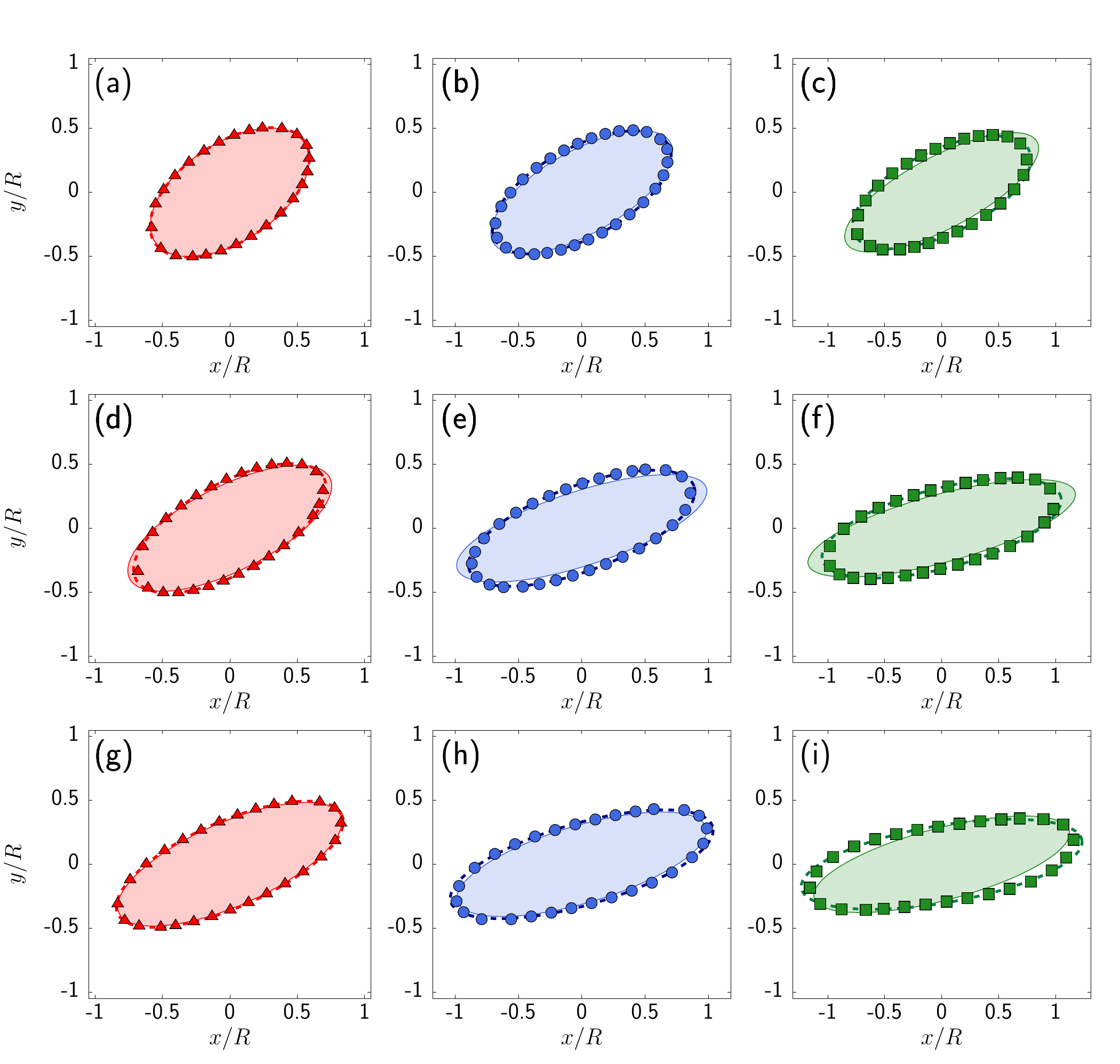}%
\caption{Droplet shapes in the shear plane for selected times indicated by arrows in Fig.~\ref{fig:transient_rk}. Numerical results for $\lambda=0.2$ (\protect \triangleRED), $\lambda=1$ (\protect \circleBLUE) $\lambda=2$ (\protect \squareGREEN) are compared with the ellipsoidal shape (dashed lines) obtained via the major/minor axes $L,B$ and inclination angle extracted from the inertia tensor in numerical simulations (see Sec.~\ref{sec:statement} for details). Colored areas indicate the ellipsoidal shapes obtained via the use of Eq.~\eqref{mmmodeladim} with $f_{1,2}=f_{1,2}^{(\mbox{\tiny EMM})}$.  }\label{fig:9shapes}%
\end{figure*}

Lastly, in Fig.~\ref{fig:transient_rk}, we report the transient values of the Taylor index for different viscosity ratios $\lambda$, comparing the EMM (Eq.~\eqref{mmmodeladim} with $f_{1,2}=f_{1,2}^{(\mbox{\tiny EMM})}$) and the original MM model (Eq.~\eqref{mmmodeladim} with $f_{1,2}=f_{1,2}^{(\mbox{\tiny MM})}$). 
When comparing model predictions and data from numerical simulations in the transient region, the EMM model matches better than the MM one. When either the viscosity ratio $\lambda$ or the capillary number $\Ca$ increase, the Taylor index $D$ predicted by both MM and EMM models shows an overshoot\footnote{Numerical simulations also show overshoots, but they are tiny and cannot be well appreciated in Fig.~\ref{fig:transient_rk}.}.  In particular, this overshoot causes the EMM model to overestimate the data of the numerical simulations during the transient dynamics; however, the steady value of the deformation predicted by the EMM model, on the contrary of the MM model, matches (by construction) the one observed in numerical simulations. 
In order to delve deeper into the origin of the mismatch observed in Fig.~\ref{fig:transient_rk} between model predictions and simulation data, we analyze in Fig.~\ref{fig:9shapes} the shape of the droplet in correspondence of some selected times for the highest $\Ca$ analyzed in Fig.~\ref{fig:transient_rk} (see arrows therein). We compare the shapes predicted by the EMM model (colored ellipses) with those coming from the numerical simulations (filled points). Dashed lines represent ellipses with both the major and minor axes ($L$ and $B$) and inclination angle $\theta$ coming from the inertia tensor computed in the numerical simulations (see Sec.~\ref{sec:methods}). Filled points and dashed lines in Fig.~\ref{fig:9shapes} show a good agreement, suggesting that the shape of the droplet remains ellipsoidal during the whole numerical simulation. 
As $\lambda$ grows, Panels (c,f,i) highlight that the rotation of the droplet induced by $\vec{\Omega}$ leads to a slight mismatch between model predictions and numerical simulations. More quantitatively, the discrepancy for $L$ and $B$ that we observe for the highest $\lambda$ in Fig.~\ref{fig:9shapes} is of the order of $10\%$ and $15\%$ in Panels (c) and (f), respectively; the error reduces to about $5\%$ in Panel (i). Concerning the inclination angle, the mismatch is of the order of $15\%$ in Panel (c), $5\%$ in Panel (f) and up to $30\%$ in Panel (i).
In the transient equiviscous case (Panel (e)), a mismatch for $L$, $B$ and $\theta$ of the same order of Panels (c and f),  is observed. 
We hasten to remark that discrepancies between numerical simulations and EMM model predictions show up only when we move close to the critical point, while being very mitigated for smaller - but still finite $\Ca$.


\section{Summary and Conclusions}\label{sec:conclusions}
We have studied the deformation process of a droplet under simple shear flow at finite capillary numbers ($\Ca \sim {\cal O}(1)$) and extracted relevant physical information to extend the popular Maffettone-Minale (MM) model~\cite{maffettone1998equation}. 
In the latter, the shape of the droplet is described by the second-order tensor $\vec{S}$, while the flow properties are expressed by the symmetric ($\vec{E}$) and asymmetric ($\vec{\Omega}$) parts of the velocity gradient $\boldsymbol\nabla\vec{u}$ (see Sec.~\ref{sec:statement}).
The MM model hinges on two parameters ($f_1(\lambda)$ and $f_2(\lambda)$) that are linked to the loading time of the droplet and the deformation induced by the flow, respectively (see Eq.~\eqref{f1f2mm}). In the MM model, such parameters are found by requiring a theoretical matching with results from perturbation theory at small $\Ca$~\cite{taylor1932viscosity,frankel1970constitutive,rallison1980note}:
for this reason, the MM model is not appropriate for a quantitative prediction of the deformation $D$ and the inclination angle $\theta$ for finite values of the capillary number $\Ca\sim {\cal O}(1)$  (see Figs.~\ref{fig:stationary_twopanels_angles}-\ref{fig:stationary_twopanels_defos}).
In this paper, we provide an extension of the MM model (referred as EMM model) via a direct evaluation of both the loading time and the steady deformation at $\Ca\sim{\cal O}(1)$ with independent numerical simulations. The latter have been performed with the Immersed Boundary - Lattice Boltzmann (IB-LB) method~\cite{book:kruger,li2019finite,guglietta2020effects}, that allowed precise characterization of the droplet dynamics from small deformations up to subcritical regimes. 
The proposed EMM model is given by: 
\begin{equation}\label{eq:emmmodeladim}
    \frac{d\vec{S}}{dt'}-\Ca\,\left(\boldsymbol{\Omega}\cdot\boldsymbol{S}-\boldsymbol{S}\cdot\boldsymbol{\Omega}\right)= -f_{1}^{(\mbox{\tiny EMM})}(\lambda,\Ca) \left[\boldsymbol{S}-g\left(\boldsymbol{S}\right)\boldsymbol{I}\right]+\Ca\,f_{2}^{(\mbox{\tiny EMM})}(\lambda,\Ca) \left(\boldsymbol{E}\cdot\vec{S}+\vec{S}\cdot\boldsymbol{E}\right) \ ,
\end{equation}
where the model coefficients $f_{1,2}^{(\mbox{\tiny EMM})}(\lambda,\Ca)$ have been characterized as a function of both $\lambda$ and $\Ca$ (see Figs.~\ref{fig:twopanels_deltas_f1},~\ref{fig:sub-crit_Lambda1},~\ref{fig:f2} and Eq.~\eqref{eq:f12EMM}). The EMM model embeds a realistic loading time as well as a realistic deformation at $\Ca \sim {\cal O}(1)$, and its predictions on the steady deformation angle are found to be in good agreement with the results of numerical simulations (see Fig.~\ref{fig:stationary_twopanels_angles}). This result is not trivial, since we are able to predict the steady deformation angle via an independent measure of the loading time; moreover, it can suggest purposeful experimental setups aimed at the evaluation of stationary state properties. Improved agreement is also observed during the whole transient process of droplet deformation (see Fig.~\ref{fig:transient_rk}), although some enhanced overshoots are observed with respect to the numerical simulations, suggesting that a further model extension is needed to better capture the full dynamics of droplet deformation. Investigating whether these extensions would be possible by retaining a second order tensorial description or by including an higher order expansion is surely an interesting topic. 
Besides, studying how the loading time is affected by additional interfacial complexities, such as interfacial viscosities and surfactants, could also lead to model refinements and adaptations for capsules or vesicles.
Also, the results in Fig.~\ref{fig:sub-crit_Lambda1} pave the way for the determination of $\Cacr$ via the EMM model, including a detailed analysis for the characterization of the droplet shapes very close to criticality. Another prospective for future research would be to investigate the degree of universality of our results with respect to a change in the topology of the driving flows. This could help, for example, to extend earlier studies on the statistics of droplet deformation and breakup in a generic time-dependent flow~\cite{biferale2014deformation,elghobashi2019direct,spandan2016deformation}, in that one can study how such statistics is affected by the interplay between the characteristic times of the driving flow and the characteristic droplet time that depends on the local flow intensity.

\begin{acknowledgments}
We wish to acknowledge Fabio Bonaccorso for support. We also acknowledge useful discussion with Luca Biferale. This work received funding from the European Research Council (ERC) under the European Union's Horizon 2020 research and innovation programme (grant agreement No 882340).

\end{acknowledgments}

\bibliography{apssamp}

\begin{thebibliography}{77}%
\makeatletter
\providecommand \@ifxundefined [1]{%
 \@ifx{#1\undefined}
}%
\providecommand \@ifnum [1]{%
 \ifnum #1\expandafter \@firstoftwo
 \else \expandafter \@secondoftwo
 \fi
}%
\providecommand \@ifx [1]{%
 \ifx #1\expandafter \@firstoftwo
 \else \expandafter \@secondoftwo
 \fi
}%
\providecommand \natexlab [1]{#1}%
\providecommand \enquote  [1]{``#1''}%
\providecommand \bibnamefont  [1]{#1}%
\providecommand \bibfnamefont [1]{#1}%
\providecommand \citenamefont [1]{#1}%
\providecommand \href@noop [0]{\@secondoftwo}%
\providecommand \href [0]{\begingroup \@sanitize@url \@href}%
\providecommand \@href[1]{\@@startlink{#1}\@@href}%
\providecommand \@@href[1]{\endgroup#1\@@endlink}%
\providecommand \@sanitize@url [0]{\catcode `\\12\catcode `\$12\catcode
  `\&12\catcode `\#12\catcode `\^12\catcode `\_12\catcode `\%12\relax}%
\providecommand \@@startlink[1]{}%
\providecommand \@@endlink[0]{}%
\providecommand \url  [0]{\begingroup\@sanitize@url \@url }%
\providecommand \@url [1]{\endgroup\@href {#1}{\urlprefix }}%
\providecommand \urlprefix  [0]{URL }%
\providecommand \Eprint [0]{\href }%
\providecommand \doibase [0]{https://doi.org/}%
\providecommand \selectlanguage [0]{\@gobble}%
\providecommand \bibinfo  [0]{\@secondoftwo}%
\providecommand \bibfield  [0]{\@secondoftwo}%
\providecommand \translation [1]{[#1]}%
\providecommand \BibitemOpen [0]{}%
\providecommand \bibitemStop [0]{}%
\providecommand \bibitemNoStop [0]{.\EOS\space}%
\providecommand \EOS [0]{\spacefactor3000\relax}%
\providecommand \BibitemShut  [1]{\csname bibitem#1\endcsname}%
\let\auto@bib@innerbib\@empty
\bibitem [{\citenamefont {Taylor}(1932)}]{taylor1932viscosity}%
  \BibitemOpen
  \bibfield  {author} {\bibinfo {author} {\bibfnamefont {G.~I.}\ \bibnamefont
  {Taylor}},\ }\bibfield  {title} {\bibinfo {title} {The viscosity of a fluid
  containing small drops of another fluid},\ }\href@noop {} {\bibfield
  {journal} {\bibinfo  {journal} {Proceedings of the Royal Society of London.
  Series A, Containing Papers of a Mathematical and Physical Character}\
  }\textbf {\bibinfo {volume} {138}},\ \bibinfo {pages} {41} (\bibinfo {year}
  {1932})}\BibitemShut {NoStop}%
\bibitem [{\citenamefont {Taylor}(1934)}]{taylor1934formation}%
  \BibitemOpen
  \bibfield  {author} {\bibinfo {author} {\bibfnamefont {G.~I.}\ \bibnamefont
  {Taylor}},\ }\bibfield  {title} {\bibinfo {title} {The formation of emulsions
  in definable fields of flow},\ }\href@noop {} {\bibfield  {journal} {\bibinfo
   {journal} {Proceedings of the Royal Society of London. Series A, Containing
  Papers of a Mathematical and Physical Character}\ }\textbf {\bibinfo {volume}
  {146}},\ \bibinfo {pages} {501} (\bibinfo {year} {1934})}\BibitemShut
  {NoStop}%
\bibitem [{\citenamefont {Einstein}(1906)}]{einstein1906new}%
  \BibitemOpen
  \bibfield  {author} {\bibinfo {author} {\bibfnamefont {A.}~\bibnamefont
  {Einstein}},\ }\bibfield  {title} {\bibinfo {title} {A new determination of
  molecular dimensions},\ }\href@noop {} {\bibfield  {journal} {\bibinfo
  {journal} {Ann. Phys.}\ }\textbf {\bibinfo {volume} {19}},\ \bibinfo {pages}
  {289} (\bibinfo {year} {1906})}\BibitemShut {NoStop}%
\bibitem [{\citenamefont {Chaffey}\ and\ \citenamefont
  {Brenner}(1967)}]{chaffey1967second}%
  \BibitemOpen
  \bibfield  {author} {\bibinfo {author} {\bibfnamefont {C.~E.}\ \bibnamefont
  {Chaffey}}\ and\ \bibinfo {author} {\bibfnamefont {H.}~\bibnamefont
  {Brenner}},\ }\bibfield  {title} {\bibinfo {title} {A second-order theory for
  shear deformation of drops},\ }\href@noop {} {\bibfield  {journal} {\bibinfo
  {journal} {Journal of Colloid and Interface Science}\ }\textbf {\bibinfo
  {volume} {24}},\ \bibinfo {pages} {258} (\bibinfo {year} {1967})}\BibitemShut
  {NoStop}%
\bibitem [{\citenamefont {Barthes-Biesel}\ and\ \citenamefont
  {Acrivos}(1973)}]{barthes1973deformation}%
  \BibitemOpen
  \bibfield  {author} {\bibinfo {author} {\bibfnamefont {D.}~\bibnamefont
  {Barthes-Biesel}}\ and\ \bibinfo {author} {\bibfnamefont {A.}~\bibnamefont
  {Acrivos}},\ }\bibfield  {title} {\bibinfo {title} {Deformation and burst of
  a liquid droplet freely suspended in a linear shear field},\ }\href@noop {}
  {\bibfield  {journal} {\bibinfo  {journal} {Journal of Fluid Mechanics}\
  }\textbf {\bibinfo {volume} {61}},\ \bibinfo {pages} {1} (\bibinfo {year}
  {1973})}\BibitemShut {NoStop}%
\bibitem [{\citenamefont {Cox}(1969)}]{cox1969deformation}%
  \BibitemOpen
  \bibfield  {author} {\bibinfo {author} {\bibfnamefont {R.}~\bibnamefont
  {Cox}},\ }\bibfield  {title} {\bibinfo {title} {The deformation of a drop in
  a general time-dependent fluid flow},\ }\href@noop {} {\bibfield  {journal}
  {\bibinfo  {journal} {Journal of Fluid Mechanics}\ }\textbf {\bibinfo
  {volume} {37}},\ \bibinfo {pages} {601} (\bibinfo {year} {1969})}\BibitemShut
  {NoStop}%
\bibitem [{\citenamefont {Frankel}\ and\ \citenamefont
  {Acrivos}(1970)}]{frankel1970constitutive}%
  \BibitemOpen
  \bibfield  {author} {\bibinfo {author} {\bibfnamefont {N.}~\bibnamefont
  {Frankel}}\ and\ \bibinfo {author} {\bibfnamefont {A.}~\bibnamefont
  {Acrivos}},\ }\bibfield  {title} {\bibinfo {title} {The constitutive equation
  for a dilute emulsion},\ }\href@noop {} {\bibfield  {journal} {\bibinfo
  {journal} {Journal of Fluid Mechanics}\ }\textbf {\bibinfo {volume} {44}},\
  \bibinfo {pages} {65} (\bibinfo {year} {1970})}\BibitemShut {NoStop}%
\bibitem [{\citenamefont {Hakimi}\ and\ \citenamefont
  {Schowalter}(1980)}]{hakimi1980effects}%
  \BibitemOpen
  \bibfield  {author} {\bibinfo {author} {\bibfnamefont {F.}~\bibnamefont
  {Hakimi}}\ and\ \bibinfo {author} {\bibfnamefont {W.}~\bibnamefont
  {Schowalter}},\ }\bibfield  {title} {\bibinfo {title} {The effects of shear
  and vorticity on deformation of a drop},\ }\href@noop {} {\bibfield
  {journal} {\bibinfo  {journal} {Journal of Fluid Mechanics}\ }\textbf
  {\bibinfo {volume} {98}},\ \bibinfo {pages} {635} (\bibinfo {year}
  {1980})}\BibitemShut {NoStop}%
\bibitem [{\citenamefont {Youngren}\ and\ \citenamefont
  {Acrivos}(1976)}]{youngren1976shape}%
  \BibitemOpen
  \bibfield  {author} {\bibinfo {author} {\bibfnamefont {G.}~\bibnamefont
  {Youngren}}\ and\ \bibinfo {author} {\bibfnamefont {A.}~\bibnamefont
  {Acrivos}},\ }\bibfield  {title} {\bibinfo {title} {On the shape of a gas
  bubble in a viscous extensional flow},\ }\href@noop {} {\bibfield  {journal}
  {\bibinfo  {journal} {Journal of Fluid Mechanics}\ }\textbf {\bibinfo
  {volume} {76}},\ \bibinfo {pages} {433} (\bibinfo {year} {1976})}\BibitemShut
  {NoStop}%
\bibitem [{\citenamefont {Greco}(2002)}]{greco2002drop}%
  \BibitemOpen
  \bibfield  {author} {\bibinfo {author} {\bibfnamefont {F.}~\bibnamefont
  {Greco}},\ }\bibfield  {title} {\bibinfo {title} {Drop deformation for
  non-newtonian fluids in slow flows},\ }\href@noop {} {\bibfield  {journal}
  {\bibinfo  {journal} {Journal of Non-Newtonian Fluid Mechanics}\ }\textbf
  {\bibinfo {volume} {107}},\ \bibinfo {pages} {111} (\bibinfo {year}
  {2002})}\BibitemShut {NoStop}%
\bibitem [{\citenamefont {Vananroye}\ \emph {et~al.}(2006)\citenamefont
  {Vananroye}, \citenamefont {Van~Puyvelde},\ and\ \citenamefont
  {Moldenaers}}]{vananroye2006effect}%
  \BibitemOpen
  \bibfield  {author} {\bibinfo {author} {\bibfnamefont {A.}~\bibnamefont
  {Vananroye}}, \bibinfo {author} {\bibfnamefont {P.}~\bibnamefont
  {Van~Puyvelde}},\ and\ \bibinfo {author} {\bibfnamefont {P.}~\bibnamefont
  {Moldenaers}},\ }\bibfield  {title} {\bibinfo {title} {Effect of confinement
  on droplet breakup in sheared emulsions},\ }\href@noop {} {\bibfield
  {journal} {\bibinfo  {journal} {Langmuir}\ }\textbf {\bibinfo {volume}
  {22}},\ \bibinfo {pages} {3972} (\bibinfo {year} {2006})}\BibitemShut
  {NoStop}%
\bibitem [{\citenamefont {Rallison}(1984)}]{rallison1984deformation}%
  \BibitemOpen
  \bibfield  {author} {\bibinfo {author} {\bibfnamefont {J.}~\bibnamefont
  {Rallison}},\ }\bibfield  {title} {\bibinfo {title} {The deformation of small
  viscous drops and bubbles in shear flows},\ }\href@noop {} {\bibfield
  {journal} {\bibinfo  {journal} {Annual Review of Fluid Mechanics}\ }\textbf
  {\bibinfo {volume} {16}},\ \bibinfo {pages} {45} (\bibinfo {year}
  {1984})}\BibitemShut {NoStop}%
\bibitem [{\citenamefont {Stone}(1994)}]{stone1994dynamics}%
  \BibitemOpen
  \bibfield  {author} {\bibinfo {author} {\bibfnamefont {H.~A.}\ \bibnamefont
  {Stone}},\ }\bibfield  {title} {\bibinfo {title} {Dynamics of drop
  deformation and breakup in viscous fluids},\ }\href@noop {} {\bibfield
  {journal} {\bibinfo  {journal} {Annual Review of Fluid Mechanics}\ }\textbf
  {\bibinfo {volume} {26}},\ \bibinfo {pages} {65} (\bibinfo {year}
  {1994})}\BibitemShut {NoStop}%
\bibitem [{\citenamefont {Fischer}\ and\ \citenamefont
  {Erni}(2007)}]{fischer2007emulsion}%
  \BibitemOpen
  \bibfield  {author} {\bibinfo {author} {\bibfnamefont {P.}~\bibnamefont
  {Fischer}}\ and\ \bibinfo {author} {\bibfnamefont {P.}~\bibnamefont {Erni}},\
  }\bibfield  {title} {\bibinfo {title} {Emulsion drops in external flow
  fields—the role of liquid interfaces},\ }\href@noop {} {\bibfield
  {journal} {\bibinfo  {journal} {Current Opinion in Colloid \& Interface
  Science}\ }\textbf {\bibinfo {volume} {12}},\ \bibinfo {pages} {196}
  (\bibinfo {year} {2007})}\BibitemShut {NoStop}%
\bibitem [{\citenamefont {Cristini}\ and\ \citenamefont
  {Tan}(2004)}]{cristini2004theory}%
  \BibitemOpen
  \bibfield  {author} {\bibinfo {author} {\bibfnamefont {V.}~\bibnamefont
  {Cristini}}\ and\ \bibinfo {author} {\bibfnamefont {Y.-C.}\ \bibnamefont
  {Tan}},\ }\bibfield  {title} {\bibinfo {title} {Theory and numerical
  simulation of droplet dynamics in complex flows—a review},\ }\href@noop {}
  {\bibfield  {journal} {\bibinfo  {journal} {Lab on a Chip}\ }\textbf
  {\bibinfo {volume} {4}},\ \bibinfo {pages} {257} (\bibinfo {year}
  {2004})}\BibitemShut {NoStop}%
\bibitem [{\citenamefont {Maffettone}\ and\ \citenamefont
  {Minale}(1998)}]{maffettone1998equation}%
  \BibitemOpen
  \bibfield  {author} {\bibinfo {author} {\bibfnamefont {P.}~\bibnamefont
  {Maffettone}}\ and\ \bibinfo {author} {\bibfnamefont {M.}~\bibnamefont
  {Minale}},\ }\bibfield  {title} {\bibinfo {title} {Equation of change for
  ellipsoidal drops in viscous flow},\ }\href@noop {} {\bibfield  {journal}
  {\bibinfo  {journal} {Journal of Non-Newtonian Fluid Mechanics}\ }\textbf
  {\bibinfo {volume} {78}},\ \bibinfo {pages} {227} (\bibinfo {year}
  {1998})}\BibitemShut {NoStop}%
\bibitem [{\citenamefont {Rallison}(1980)}]{rallison1980note}%
  \BibitemOpen
  \bibfield  {author} {\bibinfo {author} {\bibfnamefont {J.}~\bibnamefont
  {Rallison}},\ }\bibfield  {title} {\bibinfo {title} {Note on the
  time-dependent deformation of a viscous drop which is almost spherical},\
  }\href@noop {} {\bibfield  {journal} {\bibinfo  {journal} {Journal of Fluid
  Mechanics}\ }\textbf {\bibinfo {volume} {98}},\ \bibinfo {pages} {625}
  (\bibinfo {year} {1980})}\BibitemShut {NoStop}%
\bibitem [{\citenamefont {Torza}\ \emph {et~al.}(1972)\citenamefont {Torza},
  \citenamefont {Cox},\ and\ \citenamefont {Mason}}]{torza1972particle}%
  \BibitemOpen
  \bibfield  {author} {\bibinfo {author} {\bibfnamefont {S.}~\bibnamefont
  {Torza}}, \bibinfo {author} {\bibfnamefont {R.}~\bibnamefont {Cox}},\ and\
  \bibinfo {author} {\bibfnamefont {S.}~\bibnamefont {Mason}},\ }\bibfield
  {title} {\bibinfo {title} {Particle motions in sheared suspensions xxvii.
  transient and steady deformation and burst of liquid drops},\ }\href@noop {}
  {\bibfield  {journal} {\bibinfo  {journal} {Journal of Colloid and Interface
  Science}\ }\textbf {\bibinfo {volume} {38}},\ \bibinfo {pages} {395}
  (\bibinfo {year} {1972})}\BibitemShut {NoStop}%
\bibitem [{\citenamefont {Guido}\ and\ \citenamefont
  {Villone}(1998)}]{guido1998three}%
  \BibitemOpen
  \bibfield  {author} {\bibinfo {author} {\bibfnamefont {S.}~\bibnamefont
  {Guido}}\ and\ \bibinfo {author} {\bibfnamefont {M.}~\bibnamefont
  {Villone}},\ }\bibfield  {title} {\bibinfo {title} {Three-dimensional shape
  of a drop under simple shear flow},\ }\href@noop {} {\bibfield  {journal}
  {\bibinfo  {journal} {Journal of Rheology}\ }\textbf {\bibinfo {volume}
  {42}},\ \bibinfo {pages} {395} (\bibinfo {year} {1998})}\BibitemShut
  {NoStop}%
\bibitem [{\citenamefont {Bentley}\ and\ \citenamefont
  {Leal}(1986)}]{bentley1986experimental}%
  \BibitemOpen
  \bibfield  {author} {\bibinfo {author} {\bibfnamefont {B.}~\bibnamefont
  {Bentley}}\ and\ \bibinfo {author} {\bibfnamefont {L.~G.}\ \bibnamefont
  {Leal}},\ }\bibfield  {title} {\bibinfo {title} {An experimental
  investigation of drop deformation and breakup in steady, two-dimensional
  linear flows},\ }\href@noop {} {\bibfield  {journal} {\bibinfo  {journal}
  {Journal of Fluid Mechanics}\ }\textbf {\bibinfo {volume} {167}},\ \bibinfo
  {pages} {241} (\bibinfo {year} {1986})}\BibitemShut {NoStop}%
\bibitem [{\citenamefont {Maffettone}\ and\ \citenamefont
  {Greco}(2004)}]{maffettone2004ellipsoidal}%
  \BibitemOpen
  \bibfield  {author} {\bibinfo {author} {\bibfnamefont {P.~L.}\ \bibnamefont
  {Maffettone}}\ and\ \bibinfo {author} {\bibfnamefont {F.}~\bibnamefont
  {Greco}},\ }\bibfield  {title} {\bibinfo {title} {Ellipsoidal drop model for
  single drop dynamics with non-newtonian fluids},\ }\href@noop {} {\bibfield
  {journal} {\bibinfo  {journal} {Journal of Rheology}\ }\textbf {\bibinfo
  {volume} {48}},\ \bibinfo {pages} {83} (\bibinfo {year} {2004})}\BibitemShut
  {NoStop}%
\bibitem [{\citenamefont {Minale}(2004)}]{minale2004deformation}%
  \BibitemOpen
  \bibfield  {author} {\bibinfo {author} {\bibfnamefont {M.}~\bibnamefont
  {Minale}},\ }\bibfield  {title} {\bibinfo {title} {Deformation of a
  non-newtonian ellipsoidal drop in a non-newtonian matrix: extension of
  maffettone--minale model},\ }\href@noop {} {\bibfield  {journal} {\bibinfo
  {journal} {Journal of Non-Newtonian Fluid Mechanics}\ }\textbf {\bibinfo
  {volume} {123}},\ \bibinfo {pages} {151} (\bibinfo {year}
  {2004})}\BibitemShut {NoStop}%
\bibitem [{\citenamefont {Guido}\ \emph {et~al.}(2003)\citenamefont {Guido},
  \citenamefont {Simeone},\ and\ \citenamefont {Greco}}]{guido2003deformation}%
  \BibitemOpen
  \bibfield  {author} {\bibinfo {author} {\bibfnamefont {S.}~\bibnamefont
  {Guido}}, \bibinfo {author} {\bibfnamefont {M.}~\bibnamefont {Simeone}},\
  and\ \bibinfo {author} {\bibfnamefont {F.}~\bibnamefont {Greco}},\ }\bibfield
   {title} {\bibinfo {title} {Deformation of a newtonian drop in a viscoelastic
  matrix under steady shear flow: experimental validation of slow flow
  theory},\ }\href@noop {} {\bibfield  {journal} {\bibinfo  {journal} {Journal
  of Non-Newtonian Fluid Mechanics}\ }\textbf {\bibinfo {volume} {114}},\
  \bibinfo {pages} {65} (\bibinfo {year} {2003})}\BibitemShut {NoStop}%
\bibitem [{\citenamefont {Minale}(2008)}]{minale2008phenomenological}%
  \BibitemOpen
  \bibfield  {author} {\bibinfo {author} {\bibfnamefont {M.}~\bibnamefont
  {Minale}},\ }\bibfield  {title} {\bibinfo {title} {A phenomenological model
  for wall effects on the deformation of an ellipsoidal drop in viscous flow},\
  }\href@noop {} {\bibfield  {journal} {\bibinfo  {journal} {Rheologica acta}\
  }\textbf {\bibinfo {volume} {47}},\ \bibinfo {pages} {667} (\bibinfo {year}
  {2008})}\BibitemShut {NoStop}%
\bibitem [{\citenamefont {Minale}\ \emph {et~al.}(2010)\citenamefont {Minale},
  \citenamefont {Caserta},\ and\ \citenamefont
  {Guido}}]{minale2010microconfined}%
  \BibitemOpen
  \bibfield  {author} {\bibinfo {author} {\bibfnamefont {M.}~\bibnamefont
  {Minale}}, \bibinfo {author} {\bibfnamefont {S.}~\bibnamefont {Caserta}},\
  and\ \bibinfo {author} {\bibfnamefont {S.}~\bibnamefont {Guido}},\ }\bibfield
   {title} {\bibinfo {title} {Microconfined shear deformation of a droplet in
  an equiviscous non-newtonian immiscible fluid: Experiments and modeling},\
  }\href@noop {} {\bibfield  {journal} {\bibinfo  {journal} {Langmuir}\
  }\textbf {\bibinfo {volume} {26}},\ \bibinfo {pages} {126} (\bibinfo {year}
  {2010})}\BibitemShut {NoStop}%
\bibitem [{\citenamefont {Arora}\ \emph {et~al.}(2004)\citenamefont {Arora},
  \citenamefont {Behr},\ and\ \citenamefont {Pasquali}}]{art:arora04}%
  \BibitemOpen
  \bibfield  {author} {\bibinfo {author} {\bibfnamefont {D.}~\bibnamefont
  {Arora}}, \bibinfo {author} {\bibfnamefont {M.}~\bibnamefont {Behr}},\ and\
  \bibinfo {author} {\bibfnamefont {M.}~\bibnamefont {Pasquali}},\ }\bibfield
  {title} {\bibinfo {title} {A tensor-based measure for estimating blood
  damage.},\ }\href@noop {} {\bibfield  {journal} {\bibinfo  {journal}
  {Artificial organs}\ }\textbf {\bibinfo {volume} {28 11}},\ \bibinfo {pages}
  {1002} (\bibinfo {year} {2004})}\BibitemShut {NoStop}%
\bibitem [{\citenamefont {Grace}(1982)}]{grace1982dispersion}%
  \BibitemOpen
  \bibfield  {author} {\bibinfo {author} {\bibfnamefont {H.~P.}\ \bibnamefont
  {Grace}},\ }\bibfield  {title} {\bibinfo {title} {Dispersion phenomena in
  high viscosity immiscible fluid systems and application of static mixers as
  dispersion devices in such systems},\ }\href@noop {} {\bibfield  {journal}
  {\bibinfo  {journal} {Chemical Engineering Communications}\ }\textbf
  {\bibinfo {volume} {14}},\ \bibinfo {pages} {225} (\bibinfo {year}
  {1982})}\BibitemShut {NoStop}%
\bibitem [{\citenamefont {Almusallam}\ \emph {et~al.}(2000)\citenamefont
  {Almusallam}, \citenamefont {Larson},\ and\ \citenamefont
  {Solomon}}]{almusallam2000constitutive}%
  \BibitemOpen
  \bibfield  {author} {\bibinfo {author} {\bibfnamefont {A.~S.}\ \bibnamefont
  {Almusallam}}, \bibinfo {author} {\bibfnamefont {R.~G.}\ \bibnamefont
  {Larson}},\ and\ \bibinfo {author} {\bibfnamefont {M.~J.}\ \bibnamefont
  {Solomon}},\ }\bibfield  {title} {\bibinfo {title} {A constitutive model for
  the prediction of ellipsoidal droplet shapes and stresses in immiscible
  blends},\ }\href@noop {} {\bibfield  {journal} {\bibinfo  {journal} {Journal
  of Rheology}\ }\textbf {\bibinfo {volume} {44}},\ \bibinfo {pages} {1055}
  (\bibinfo {year} {2000})}\BibitemShut {NoStop}%
\bibitem [{\citenamefont {Jackson}\ and\ \citenamefont
  {Tucker~III}(2003)}]{jackson2003model}%
  \BibitemOpen
  \bibfield  {author} {\bibinfo {author} {\bibfnamefont {N.~E.}\ \bibnamefont
  {Jackson}}\ and\ \bibinfo {author} {\bibfnamefont {C.~L.}\ \bibnamefont
  {Tucker~III}},\ }\bibfield  {title} {\bibinfo {title} {A model for large
  deformation of an ellipsoidal droplet with interfacial tension},\ }\href@noop
  {} {\bibfield  {journal} {\bibinfo  {journal} {Journal of Rheology}\ }\textbf
  {\bibinfo {volume} {47}},\ \bibinfo {pages} {659} (\bibinfo {year}
  {2003})}\BibitemShut {NoStop}%
\bibitem [{\citenamefont {Yu}\ and\ \citenamefont
  {Bousmina}(2003)}]{yu2003ellipsoidal}%
  \BibitemOpen
  \bibfield  {author} {\bibinfo {author} {\bibfnamefont {W.}~\bibnamefont
  {Yu}}\ and\ \bibinfo {author} {\bibfnamefont {M.}~\bibnamefont {Bousmina}},\
  }\bibfield  {title} {\bibinfo {title} {Ellipsoidal model for droplet
  deformation in emulsions},\ }\href@noop {} {\bibfield  {journal} {\bibinfo
  {journal} {Journal of Rheology}\ }\textbf {\bibinfo {volume} {47}},\ \bibinfo
  {pages} {1011} (\bibinfo {year} {2003})}\BibitemShut {NoStop}%
\bibitem [{\citenamefont {Minale}(2010)}]{minale2010models}%
  \BibitemOpen
  \bibfield  {author} {\bibinfo {author} {\bibfnamefont {M.}~\bibnamefont
  {Minale}},\ }\bibfield  {title} {\bibinfo {title} {Models for the deformation
  of a single ellipsoidal drop: a review},\ }\href@noop {} {\bibfield
  {journal} {\bibinfo  {journal} {Rheologica acta}\ }\textbf {\bibinfo {volume}
  {49}},\ \bibinfo {pages} {789} (\bibinfo {year} {2010})}\BibitemShut
  {NoStop}%
\bibitem [{\citenamefont {Cristini}\ \emph {et~al.}(2003)\citenamefont
  {Cristini}, \citenamefont {Guido}, \citenamefont {Alfani}, \citenamefont
  {B{\l}awzdziewicz},\ and\ \citenamefont {Loewenberg}}]{cristini2003drop}%
  \BibitemOpen
  \bibfield  {author} {\bibinfo {author} {\bibfnamefont {V.}~\bibnamefont
  {Cristini}}, \bibinfo {author} {\bibfnamefont {S.}~\bibnamefont {Guido}},
  \bibinfo {author} {\bibfnamefont {A.}~\bibnamefont {Alfani}}, \bibinfo
  {author} {\bibfnamefont {J.}~\bibnamefont {B{\l}awzdziewicz}},\ and\ \bibinfo
  {author} {\bibfnamefont {M.}~\bibnamefont {Loewenberg}},\ }\bibfield  {title}
  {\bibinfo {title} {Drop breakup and fragment size distribution in shear
  flow},\ }\href@noop {} {\bibfield  {journal} {\bibinfo  {journal} {Journal of
  Rheology}\ }\textbf {\bibinfo {volume} {47}},\ \bibinfo {pages} {1283}
  (\bibinfo {year} {2003})}\BibitemShut {NoStop}%
\bibitem [{\citenamefont {Feng}\ and\ \citenamefont
  {Michaelides}(2004)}]{feng2004immersed}%
  \BibitemOpen
  \bibfield  {author} {\bibinfo {author} {\bibfnamefont {Z.-G.}\ \bibnamefont
  {Feng}}\ and\ \bibinfo {author} {\bibfnamefont {E.~E.}\ \bibnamefont
  {Michaelides}},\ }\bibfield  {title} {\bibinfo {title} {The immersed
  boundary-lattice boltzmann method for solving fluid--particles interaction
  problems},\ }\href@noop {} {\bibfield  {journal} {\bibinfo  {journal}
  {Journal of Computational Physics}\ }\textbf {\bibinfo {volume} {195}},\
  \bibinfo {pages} {602} (\bibinfo {year} {2004})}\BibitemShut {NoStop}%
\bibitem [{\citenamefont {Zhang}\ \emph {et~al.}(2007)\citenamefont {Zhang},
  \citenamefont {Johnson},\ and\ \citenamefont
  {Popel}}]{zhangImmersedBoundaryLattice2007}%
  \BibitemOpen
  \bibfield  {author} {\bibinfo {author} {\bibfnamefont {J.}~\bibnamefont
  {Zhang}}, \bibinfo {author} {\bibfnamefont {P.~C.}\ \bibnamefont {Johnson}},\
  and\ \bibinfo {author} {\bibfnamefont {A.~S.}\ \bibnamefont {Popel}},\
  }\bibfield  {title} {\bibinfo {title} {An immersed boundary lattice
  {{Boltzmann}} approach to simulate deformable liquid capsules and its
  application to microscopic blood flows},\ }\href@noop {} {\bibfield
  {journal} {\bibinfo  {journal} {Physical biology}\ }\textbf {\bibinfo
  {volume} {4}},\ \bibinfo {pages} {285} (\bibinfo {year} {2007})}\BibitemShut
  {NoStop}%
\bibitem [{\citenamefont {Dupin}\ \emph {et~al.}(2007)\citenamefont {Dupin},
  \citenamefont {Halliday}, \citenamefont {Care}, \citenamefont {Alboul},\ and\
  \citenamefont {Munn}}]{dupin2007modeling}%
  \BibitemOpen
  \bibfield  {author} {\bibinfo {author} {\bibfnamefont {M.~M.}\ \bibnamefont
  {Dupin}}, \bibinfo {author} {\bibfnamefont {I.}~\bibnamefont {Halliday}},
  \bibinfo {author} {\bibfnamefont {C.~M.}\ \bibnamefont {Care}}, \bibinfo
  {author} {\bibfnamefont {L.}~\bibnamefont {Alboul}},\ and\ \bibinfo {author}
  {\bibfnamefont {L.~L.}\ \bibnamefont {Munn}},\ }\bibfield  {title} {\bibinfo
  {title} {Modeling the flow of dense suspensions of deformable particles in
  three dimensions},\ }\href@noop {} {\bibfield  {journal} {\bibinfo  {journal}
  {Physical Review E}\ }\textbf {\bibinfo {volume} {75}},\ \bibinfo {pages}
  {066707} (\bibinfo {year} {2007})}\BibitemShut {NoStop}%
\bibitem [{\citenamefont {Krüger}\ \emph {et~al.}(2016)\citenamefont
  {Krüger}, \citenamefont {Kusumaatmaja}, \citenamefont {Kuzmin},
  \citenamefont {Shardt}, \citenamefont {Silva},\ and\ \citenamefont
  {Viggen}}]{book:kruger}%
  \BibitemOpen
  \bibfield  {author} {\bibinfo {author} {\bibfnamefont {T.}~\bibnamefont
  {Krüger}}, \bibinfo {author} {\bibfnamefont {H.}~\bibnamefont
  {Kusumaatmaja}}, \bibinfo {author} {\bibfnamefont {A.}~\bibnamefont
  {Kuzmin}}, \bibinfo {author} {\bibfnamefont {O.}~\bibnamefont {Shardt}},
  \bibinfo {author} {\bibfnamefont {G.}~\bibnamefont {Silva}},\ and\ \bibinfo
  {author} {\bibfnamefont {E.~M.}\ \bibnamefont {Viggen}},\ }\href@noop {}
  {\emph {\bibinfo {title} {The Lattice Boltzmann Method - Principles and
  Practice}}}\ (\bibinfo {year} {2016})\BibitemShut {NoStop}%
\bibitem [{\citenamefont {Kr\"{u}ger}(2012)}]{thesis:kruger}%
  \BibitemOpen
  \bibfield  {author} {\bibinfo {author} {\bibfnamefont {T.}~\bibnamefont
  {Kr\"{u}ger}},\ }\emph {\bibinfo {title} {Computer Simulation Study of
  Collective Phenomena in Dense Suspensions of Red Blood Cells under Shear}},\
  \href@noop {} {Ph.D. thesis} (\bibinfo {year} {2012})\BibitemShut {NoStop}%
\bibitem [{\citenamefont {Li}\ and\ \citenamefont
  {Zhang}(2019)}]{li2019finite}%
  \BibitemOpen
  \bibfield  {author} {\bibinfo {author} {\bibfnamefont {P.}~\bibnamefont
  {Li}}\ and\ \bibinfo {author} {\bibfnamefont {J.}~\bibnamefont {Zhang}},\
  }\bibfield  {title} {\bibinfo {title} {A finite difference method with
  subsampling for immersed boundary simulations of the capsule dynamics with
  viscoelastic membranes},\ }\href@noop {} {\bibfield  {journal} {\bibinfo
  {journal} {International Journal for Numerical Methods in Biomedical
  Engineering}\ }\textbf {\bibinfo {volume} {35}},\ \bibinfo {pages} {e3200}
  (\bibinfo {year} {2019})}\BibitemShut {NoStop}%
\bibitem [{\citenamefont {Guglietta}\ \emph {et~al.}(2020)\citenamefont
  {Guglietta}, \citenamefont {Behr}, \citenamefont {Biferale}, \citenamefont
  {Falcucci},\ and\ \citenamefont {Sbragaglia}}]{guglietta2020effects}%
  \BibitemOpen
  \bibfield  {author} {\bibinfo {author} {\bibfnamefont {F.}~\bibnamefont
  {Guglietta}}, \bibinfo {author} {\bibfnamefont {M.}~\bibnamefont {Behr}},
  \bibinfo {author} {\bibfnamefont {L.}~\bibnamefont {Biferale}}, \bibinfo
  {author} {\bibfnamefont {G.}~\bibnamefont {Falcucci}},\ and\ \bibinfo
  {author} {\bibfnamefont {M.}~\bibnamefont {Sbragaglia}},\ }\bibfield  {title}
  {\bibinfo {title} {On the effects of membrane viscosity on transient red
  blood cell dynamics},\ }\href@noop {} {\bibfield  {journal} {\bibinfo
  {journal} {Soft Matter}\ }\textbf {\bibinfo {volume} {16}},\ \bibinfo {pages}
  {6191} (\bibinfo {year} {2020})}\BibitemShut {NoStop}%
\bibitem [{\citenamefont {Li}\ and\ \citenamefont
  {Zhang}(2021)}]{li2021similar}%
  \BibitemOpen
  \bibfield  {author} {\bibinfo {author} {\bibfnamefont {P.}~\bibnamefont
  {Li}}\ and\ \bibinfo {author} {\bibfnamefont {J.}~\bibnamefont {Zhang}},\
  }\bibfield  {title} {\bibinfo {title} {Similar but distinct roles of membrane
  and interior fluid viscosities in capsule dynamics in shear flows},\
  }\href@noop {} {\bibfield  {journal} {\bibinfo  {journal} {Cardiovascular
  Engineering and Technology}\ }\textbf {\bibinfo {volume} {12}},\ \bibinfo
  {pages} {232} (\bibinfo {year} {2021})}\BibitemShut {NoStop}%
\bibitem [{\citenamefont {Kr{\"u}ger}\ \emph {et~al.}(2013)\citenamefont
  {Kr{\"u}ger}, \citenamefont {Gross}, \citenamefont {Raabe},\ and\
  \citenamefont {Varnik}}]{kruger2013crossover}%
  \BibitemOpen
  \bibfield  {author} {\bibinfo {author} {\bibfnamefont {T.}~\bibnamefont
  {Kr{\"u}ger}}, \bibinfo {author} {\bibfnamefont {M.}~\bibnamefont {Gross}},
  \bibinfo {author} {\bibfnamefont {D.}~\bibnamefont {Raabe}},\ and\ \bibinfo
  {author} {\bibfnamefont {F.}~\bibnamefont {Varnik}},\ }\bibfield  {title}
  {\bibinfo {title} {Crossover from tumbling to tank-treading-like motion in
  dense simulated suspensions of red blood cells},\ }\href@noop {} {\bibfield
  {journal} {\bibinfo  {journal} {Soft Matter}\ }\textbf {\bibinfo {volume}
  {9}},\ \bibinfo {pages} {9008} (\bibinfo {year} {2013})}\BibitemShut
  {NoStop}%
\bibitem [{\citenamefont {Kr{\"u}ger}\ \emph {et~al.}(2014)\citenamefont
  {Kr{\"u}ger}, \citenamefont {Holmes},\ and\ \citenamefont
  {Coveney}}]{art:kruger14deformability}%
  \BibitemOpen
  \bibfield  {author} {\bibinfo {author} {\bibfnamefont {T.}~\bibnamefont
  {Kr{\"u}ger}}, \bibinfo {author} {\bibfnamefont {D.}~\bibnamefont {Holmes}},\
  and\ \bibinfo {author} {\bibfnamefont {P.~V.}\ \bibnamefont {Coveney}},\
  }\bibfield  {title} {\bibinfo {title} {Deformability-based red blood cell
  separation in deterministic lateral displacement devices—a simulation
  study},\ }\href@noop {} {\bibfield  {journal} {\bibinfo  {journal}
  {Biomicrofluidics}\ }\textbf {\bibinfo {volume} {8}},\ \bibinfo {pages}
  {054114} (\bibinfo {year} {2014})}\BibitemShut {NoStop}%
\bibitem [{\citenamefont {Gekle}(2016)}]{gekle2016strongly}%
  \BibitemOpen
  \bibfield  {author} {\bibinfo {author} {\bibfnamefont {S.}~\bibnamefont
  {Gekle}},\ }\bibfield  {title} {\bibinfo {title} {Strongly accelerated
  margination of active particles in blood flow},\ }\href@noop {} {\bibfield
  {journal} {\bibinfo  {journal} {Biophysical journal}\ }\textbf {\bibinfo
  {volume} {110}},\ \bibinfo {pages} {514} (\bibinfo {year}
  {2016})}\BibitemShut {NoStop}%
\bibitem [{\citenamefont {Rallison}\ and\ \citenamefont
  {Acrivos}(1978)}]{rallison1978numerical}%
  \BibitemOpen
  \bibfield  {author} {\bibinfo {author} {\bibfnamefont {J.}~\bibnamefont
  {Rallison}}\ and\ \bibinfo {author} {\bibfnamefont {A.}~\bibnamefont
  {Acrivos}},\ }\bibfield  {title} {\bibinfo {title} {A numerical study of the
  deformation and burst of a viscous drop in an extensional flow},\ }\href@noop
  {} {\bibfield  {journal} {\bibinfo  {journal} {Journal of Fluid Mechanics}\
  }\textbf {\bibinfo {volume} {89}},\ \bibinfo {pages} {191} (\bibinfo {year}
  {1978})}\BibitemShut {NoStop}%
\bibitem [{\citenamefont {Pozrikidis}\ \emph {et~al.}(1992)\citenamefont
  {Pozrikidis} \emph {et~al.}}]{pozrikidis1992boundary}%
  \BibitemOpen
  \bibfield  {author} {\bibinfo {author} {\bibfnamefont {C.}~\bibnamefont
  {Pozrikidis}} \emph {et~al.},\ }\href@noop {} {\emph {\bibinfo {title}
  {Boundary integral and singularity methods for linearized viscous flow}}}\
  (\bibinfo  {publisher} {Cambridge university press},\ \bibinfo {year}
  {1992})\BibitemShut {NoStop}%
\bibitem [{\citenamefont {Gounley}\ \emph {et~al.}(2016)\citenamefont
  {Gounley}, \citenamefont {Boedec}, \citenamefont {Jaeger},\ and\
  \citenamefont {Leonetti}}]{art:gounley16}%
  \BibitemOpen
  \bibfield  {author} {\bibinfo {author} {\bibfnamefont {J.}~\bibnamefont
  {Gounley}}, \bibinfo {author} {\bibfnamefont {G.}~\bibnamefont {Boedec}},
  \bibinfo {author} {\bibfnamefont {M.}~\bibnamefont {Jaeger}},\ and\ \bibinfo
  {author} {\bibfnamefont {M.}~\bibnamefont {Leonetti}},\ }\bibfield  {title}
  {\bibinfo {title} {Influence of surface viscosity on droplets in shear
  flow},\ }\href@noop {} {\bibfield  {journal} {\bibinfo  {journal} {Journal of
  Fluid Mechanics}\ }\textbf {\bibinfo {volume} {791}},\ \bibinfo {pages}
  {464–494} (\bibinfo {year} {2016})}\BibitemShut {NoStop}%
\bibitem [{\citenamefont {Li}\ \emph {et~al.}(2000)\citenamefont {Li},
  \citenamefont {Renardy},\ and\ \citenamefont {Renardy}}]{li2000numerical}%
  \BibitemOpen
  \bibfield  {author} {\bibinfo {author} {\bibfnamefont {J.}~\bibnamefont
  {Li}}, \bibinfo {author} {\bibfnamefont {Y.~Y.}\ \bibnamefont {Renardy}},\
  and\ \bibinfo {author} {\bibfnamefont {M.}~\bibnamefont {Renardy}},\
  }\bibfield  {title} {\bibinfo {title} {Numerical simulation of breakup of a
  viscous drop in simple shear flow through a volume-of-fluid method},\
  }\href@noop {} {\bibfield  {journal} {\bibinfo  {journal} {Physics of
  Fluids}\ }\textbf {\bibinfo {volume} {12}},\ \bibinfo {pages} {269} (\bibinfo
  {year} {2000})}\BibitemShut {NoStop}%
\bibitem [{\citenamefont {Shan}\ and\ \citenamefont
  {Chen}(1993)}]{shan1993lattice}%
  \BibitemOpen
  \bibfield  {author} {\bibinfo {author} {\bibfnamefont {X.}~\bibnamefont
  {Shan}}\ and\ \bibinfo {author} {\bibfnamefont {H.}~\bibnamefont {Chen}},\
  }\bibfield  {title} {\bibinfo {title} {Lattice boltzmann model for simulating
  flows with multiple phases and components},\ }\href@noop {} {\bibfield
  {journal} {\bibinfo  {journal} {Physical review E}\ }\textbf {\bibinfo
  {volume} {47}},\ \bibinfo {pages} {1815} (\bibinfo {year}
  {1993})}\BibitemShut {NoStop}%
\bibitem [{\citenamefont {Shan}\ and\ \citenamefont
  {Chen}(1994)}]{shan1994simulation}%
  \BibitemOpen
  \bibfield  {author} {\bibinfo {author} {\bibfnamefont {X.}~\bibnamefont
  {Shan}}\ and\ \bibinfo {author} {\bibfnamefont {H.}~\bibnamefont {Chen}},\
  }\bibfield  {title} {\bibinfo {title} {Simulation of nonideal gases and
  liquid-gas phase transitions by the lattice boltzmann equation},\ }\href@noop
  {} {\bibfield  {journal} {\bibinfo  {journal} {Physical Review E}\ }\textbf
  {\bibinfo {volume} {49}},\ \bibinfo {pages} {2941} (\bibinfo {year}
  {1994})}\BibitemShut {NoStop}%
\bibitem [{\citenamefont {Swift}\ \emph {et~al.}(1995)\citenamefont {Swift},
  \citenamefont {Osborn},\ and\ \citenamefont {Yeomans}}]{swift1995lattice}%
  \BibitemOpen
  \bibfield  {author} {\bibinfo {author} {\bibfnamefont {M.~R.}\ \bibnamefont
  {Swift}}, \bibinfo {author} {\bibfnamefont {W.}~\bibnamefont {Osborn}},\ and\
  \bibinfo {author} {\bibfnamefont {J.}~\bibnamefont {Yeomans}},\ }\bibfield
  {title} {\bibinfo {title} {Lattice boltzmann simulation of nonideal fluids},\
  }\href@noop {} {\bibfield  {journal} {\bibinfo  {journal} {Physical review
  letters}\ }\textbf {\bibinfo {volume} {75}},\ \bibinfo {pages} {830}
  (\bibinfo {year} {1995})}\BibitemShut {NoStop}%
\bibitem [{\citenamefont {Swift}\ \emph {et~al.}(1996)\citenamefont {Swift},
  \citenamefont {Orlandini}, \citenamefont {Osborn},\ and\ \citenamefont
  {Yeomans}}]{swift1996lattice}%
  \BibitemOpen
  \bibfield  {author} {\bibinfo {author} {\bibfnamefont {M.~R.}\ \bibnamefont
  {Swift}}, \bibinfo {author} {\bibfnamefont {E.}~\bibnamefont {Orlandini}},
  \bibinfo {author} {\bibfnamefont {W.}~\bibnamefont {Osborn}},\ and\ \bibinfo
  {author} {\bibfnamefont {J.}~\bibnamefont {Yeomans}},\ }\bibfield  {title}
  {\bibinfo {title} {Lattice boltzmann simulations of liquid-gas and binary
  fluid systems},\ }\href@noop {} {\bibfield  {journal} {\bibinfo  {journal}
  {Physical Review E}\ }\textbf {\bibinfo {volume} {54}},\ \bibinfo {pages}
  {5041} (\bibinfo {year} {1996})}\BibitemShut {NoStop}%
\bibitem [{\citenamefont {Liu}\ \emph {et~al.}(2012)\citenamefont {Liu},
  \citenamefont {Valocchi},\ and\ \citenamefont {Kang}}]{liu2012three}%
  \BibitemOpen
  \bibfield  {author} {\bibinfo {author} {\bibfnamefont {H.}~\bibnamefont
  {Liu}}, \bibinfo {author} {\bibfnamefont {A.~J.}\ \bibnamefont {Valocchi}},\
  and\ \bibinfo {author} {\bibfnamefont {Q.}~\bibnamefont {Kang}},\ }\bibfield
  {title} {\bibinfo {title} {Three-dimensional lattice boltzmann model for
  immiscible two-phase flow simulations},\ }\href@noop {} {\bibfield  {journal}
  {\bibinfo  {journal} {Physical Review E}\ }\textbf {\bibinfo {volume} {85}},\
  \bibinfo {pages} {046309} (\bibinfo {year} {2012})}\BibitemShut {NoStop}%
\bibitem [{\citenamefont {Chikatamarla}\ \emph {et~al.}(2015)\citenamefont
  {Chikatamarla}, \citenamefont {Karlin} \emph
  {et~al.}}]{chikatamarla2015entropic}%
  \BibitemOpen
  \bibfield  {author} {\bibinfo {author} {\bibfnamefont {S.}~\bibnamefont
  {Chikatamarla}}, \bibinfo {author} {\bibfnamefont {I.}~\bibnamefont
  {Karlin}}, \emph {et~al.},\ }\bibfield  {title} {\bibinfo {title} {Entropic
  lattice boltzmann method for multiphase flows},\ }\href@noop {} {\bibfield
  {journal} {\bibinfo  {journal} {Physical Review Letters}\ }\textbf {\bibinfo
  {volume} {114}},\ \bibinfo {pages} {174502} (\bibinfo {year}
  {2015})}\BibitemShut {NoStop}%
\bibitem [{\citenamefont {Milan}\ \emph {et~al.}(2020)\citenamefont {Milan},
  \citenamefont {Biferale}, \citenamefont {Sbragaglia},\ and\ \citenamefont
  {Toschi}}]{milan2020sub}%
  \BibitemOpen
  \bibfield  {author} {\bibinfo {author} {\bibfnamefont {F.}~\bibnamefont
  {Milan}}, \bibinfo {author} {\bibfnamefont {L.}~\bibnamefont {Biferale}},
  \bibinfo {author} {\bibfnamefont {M.}~\bibnamefont {Sbragaglia}},\ and\
  \bibinfo {author} {\bibfnamefont {F.}~\bibnamefont {Toschi}},\ }\bibfield
  {title} {\bibinfo {title} {Sub-kolmogorov droplet dynamics in isotropic
  turbulence using a multiscale lattice boltzmann scheme},\ }\href@noop {}
  {\bibfield  {journal} {\bibinfo  {journal} {Journal of Computational
  Science}\ }\textbf {\bibinfo {volume} {45}},\ \bibinfo {pages} {101178}
  (\bibinfo {year} {2020})}\BibitemShut {NoStop}%
\bibitem [{\citenamefont {Succi}(2018)}]{succi2018lattice}%
  \BibitemOpen
  \bibfield  {author} {\bibinfo {author} {\bibfnamefont {S.}~\bibnamefont
  {Succi}},\ }\href@noop {} {\emph {\bibinfo {title} {The lattice Boltzmann
  equation: for complex states of flowing matter}}}\ (\bibinfo  {publisher}
  {Oxford University Press},\ \bibinfo {year} {2018})\BibitemShut {NoStop}%
\bibitem [{\citenamefont {Qian}\ \emph {et~al.}(1992)\citenamefont {Qian},
  \citenamefont {d'Humi{\`e}res},\ and\ \citenamefont
  {Lallemand}}]{qian1992lattice}%
  \BibitemOpen
  \bibfield  {author} {\bibinfo {author} {\bibfnamefont {Y.-H.}\ \bibnamefont
  {Qian}}, \bibinfo {author} {\bibfnamefont {D.}~\bibnamefont
  {d'Humi{\`e}res}},\ and\ \bibinfo {author} {\bibfnamefont {P.}~\bibnamefont
  {Lallemand}},\ }\bibfield  {title} {\bibinfo {title} {Lattice bgk models for
  navier-stokes equation},\ }\href@noop {} {\bibfield  {journal} {\bibinfo
  {journal} {EPL (Europhysics Letters)}\ }\textbf {\bibinfo {volume} {17}},\
  \bibinfo {pages} {479} (\bibinfo {year} {1992})}\BibitemShut {NoStop}%
\bibitem [{\citenamefont {Guo}\ \emph {et~al.}(2002)\citenamefont {Guo},
  \citenamefont {Zheng},\ and\ \citenamefont {Shi}}]{guo2002discrete}%
  \BibitemOpen
  \bibfield  {author} {\bibinfo {author} {\bibfnamefont {Z.}~\bibnamefont
  {Guo}}, \bibinfo {author} {\bibfnamefont {C.}~\bibnamefont {Zheng}},\ and\
  \bibinfo {author} {\bibfnamefont {B.}~\bibnamefont {Shi}},\ }\bibfield
  {title} {\bibinfo {title} {Discrete lattice effects on the forcing term in
  the lattice boltzmann method},\ }\href@noop {} {\bibfield  {journal}
  {\bibinfo  {journal} {Physical review E}\ }\textbf {\bibinfo {volume} {65}},\
  \bibinfo {pages} {046308} (\bibinfo {year} {2002})}\BibitemShut {NoStop}%
\bibitem [{\citenamefont {Peskin}(2002)}]{peskin2002immersed}%
  \BibitemOpen
  \bibfield  {author} {\bibinfo {author} {\bibfnamefont {C.~S.}\ \bibnamefont
  {Peskin}},\ }\bibfield  {title} {\bibinfo {title} {The immersed boundary
  method},\ }\href@noop {} {\bibfield  {journal} {\bibinfo  {journal} {Acta
  numerica}\ }\textbf {\bibinfo {volume} {11}},\ \bibinfo {pages} {479}
  (\bibinfo {year} {2002})}\BibitemShut {NoStop}%
\bibitem [{\citenamefont {Guido}(2011)}]{guido2011shear}%
  \BibitemOpen
  \bibfield  {author} {\bibinfo {author} {\bibfnamefont {S.}~\bibnamefont
  {Guido}},\ }\bibfield  {title} {\bibinfo {title} {Shear-induced droplet
  deformation: Effects of confined geometry and viscoelasticity},\ }\href@noop
  {} {\bibfield  {journal} {\bibinfo  {journal} {Current opinion in colloid \&
  interface science}\ }\textbf {\bibinfo {volume} {16}},\ \bibinfo {pages} {61}
  (\bibinfo {year} {2011})}\BibitemShut {NoStop}%
\bibitem [{\citenamefont {Renardy}(2008)}]{renardy2008effect}%
  \BibitemOpen
  \bibfield  {author} {\bibinfo {author} {\bibfnamefont {Y.}~\bibnamefont
  {Renardy}},\ }\bibfield  {title} {\bibinfo {title} {Effect of startup
  conditions on drop breakup under shear with inertia},\ }\href@noop {}
  {\bibfield  {journal} {\bibinfo  {journal} {International journal of
  multiphase flow}\ }\textbf {\bibinfo {volume} {34}},\ \bibinfo {pages} {1185}
  (\bibinfo {year} {2008})}\BibitemShut {NoStop}%
\bibitem [{\citenamefont {Lessard}\ and\ \citenamefont
  {DeMarco}(2000)}]{lessard2000significance}%
  \BibitemOpen
  \bibfield  {author} {\bibinfo {author} {\bibfnamefont {R.~R.}\ \bibnamefont
  {Lessard}}\ and\ \bibinfo {author} {\bibfnamefont {G.}~\bibnamefont
  {DeMarco}},\ }\bibfield  {title} {\bibinfo {title} {The significance of oil
  spill dispersants},\ }\href@noop {} {\bibfield  {journal} {\bibinfo
  {journal} {Spill Science \& Technology Bulletin}\ }\textbf {\bibinfo {volume}
  {6}},\ \bibinfo {pages} {59} (\bibinfo {year} {2000})}\BibitemShut {NoStop}%
\bibitem [{\citenamefont {Carvalho}\ \emph {et~al.}(2017)\citenamefont
  {Carvalho}, \citenamefont {Antuniassi}, \citenamefont {Chechetto},
  \citenamefont {Mota}, \citenamefont {de~Jesus},\ and\ \citenamefont
  {de~Carvalho}}]{carvalho2017viscosity}%
  \BibitemOpen
  \bibfield  {author} {\bibinfo {author} {\bibfnamefont {F.~K.}\ \bibnamefont
  {Carvalho}}, \bibinfo {author} {\bibfnamefont {U.~R.}\ \bibnamefont
  {Antuniassi}}, \bibinfo {author} {\bibfnamefont {R.~G.}\ \bibnamefont
  {Chechetto}}, \bibinfo {author} {\bibfnamefont {A.~A.~B.}\ \bibnamefont
  {Mota}}, \bibinfo {author} {\bibfnamefont {M.~G.}\ \bibnamefont {de~Jesus}},\
  and\ \bibinfo {author} {\bibfnamefont {L.~R.}\ \bibnamefont {de~Carvalho}},\
  }\bibfield  {title} {\bibinfo {title} {Viscosity, surface tension and droplet
  size of sprays of different formulations of insecticides and fungicides},\
  }\href@noop {} {\bibfield  {journal} {\bibinfo  {journal} {Crop Protection}\
  }\textbf {\bibinfo {volume} {101}},\ \bibinfo {pages} {19} (\bibinfo {year}
  {2017})}\BibitemShut {NoStop}%
\bibitem [{\citenamefont {Phillips}(1996)}]{phillips1996stretched}%
  \BibitemOpen
  \bibfield  {author} {\bibinfo {author} {\bibfnamefont {J.}~\bibnamefont
  {Phillips}},\ }\bibfield  {title} {\bibinfo {title} {Stretched exponential
  relaxation in molecular and electronic glasses},\ }\href@noop {} {\bibfield
  {journal} {\bibinfo  {journal} {Reports on Progress in Physics}\ }\textbf
  {\bibinfo {volume} {59}},\ \bibinfo {pages} {1133} (\bibinfo {year}
  {1996})}\BibitemShut {NoStop}%
\bibitem [{\citenamefont {Fedosov}(2010)}]{fedosov2010multiscale}%
  \BibitemOpen
  \bibfield  {author} {\bibinfo {author} {\bibfnamefont {D.~A.}\ \bibnamefont
  {Fedosov}},\ }\href@noop {} {\emph {\bibinfo {title} {Multiscale modeling of
  blood flow and soft matter}}}\ (\bibinfo  {publisher} {Citeseer},\ \bibinfo
  {year} {2010})\BibitemShut {NoStop}%
\bibitem [{\citenamefont {Palmer}\ \emph {et~al.}(1984)\citenamefont {Palmer},
  \citenamefont {Stein}, \citenamefont {Abrahams},\ and\ \citenamefont
  {Anderson}}]{palmer1984models}%
  \BibitemOpen
  \bibfield  {author} {\bibinfo {author} {\bibfnamefont {R.~G.}\ \bibnamefont
  {Palmer}}, \bibinfo {author} {\bibfnamefont {D.~L.}\ \bibnamefont {Stein}},
  \bibinfo {author} {\bibfnamefont {E.}~\bibnamefont {Abrahams}},\ and\
  \bibinfo {author} {\bibfnamefont {P.~W.}\ \bibnamefont {Anderson}},\
  }\bibfield  {title} {\bibinfo {title} {Models of hierarchically constrained
  dynamics for glassy relaxation},\ }\href@noop {} {\bibfield  {journal}
  {\bibinfo  {journal} {Physical Review Letters}\ }\textbf {\bibinfo {volume}
  {53}},\ \bibinfo {pages} {958} (\bibinfo {year} {1984})}\BibitemShut
  {NoStop}%
\bibitem [{\citenamefont {Potuzak}\ \emph {et~al.}(2011)\citenamefont
  {Potuzak}, \citenamefont {Welch},\ and\ \citenamefont
  {Mauro}}]{potuzak2011topological}%
  \BibitemOpen
  \bibfield  {author} {\bibinfo {author} {\bibfnamefont {M.}~\bibnamefont
  {Potuzak}}, \bibinfo {author} {\bibfnamefont {R.~C.}\ \bibnamefont {Welch}},\
  and\ \bibinfo {author} {\bibfnamefont {J.~C.}\ \bibnamefont {Mauro}},\
  }\bibfield  {title} {\bibinfo {title} {Topological origin of stretched
  exponential relaxation in glass},\ }\href@noop {} {\bibfield  {journal}
  {\bibinfo  {journal} {The Journal of Chemical Physics}\ }\textbf {\bibinfo
  {volume} {135}},\ \bibinfo {pages} {214502} (\bibinfo {year}
  {2011})}\BibitemShut {NoStop}%
\bibitem [{\citenamefont {Jurlewicz}\ and\ \citenamefont
  {Weron}(1993)}]{jurlewicz1993relationship}%
  \BibitemOpen
  \bibfield  {author} {\bibinfo {author} {\bibfnamefont {A.}~\bibnamefont
  {Jurlewicz}}\ and\ \bibinfo {author} {\bibfnamefont {K.}~\bibnamefont
  {Weron}},\ }\bibfield  {title} {\bibinfo {title} {A relationship between
  asymmetric l{\'e}vy-stable distributions and the dielectric susceptibility},\
  }\href@noop {} {\bibfield  {journal} {\bibinfo  {journal} {Journal of
  Statistical Physics}\ }\textbf {\bibinfo {volume} {73}},\ \bibinfo {pages}
  {69} (\bibinfo {year} {1993})}\BibitemShut {NoStop}%
\bibitem [{\citenamefont {Elton}(2018)}]{elton2018stretched}%
  \BibitemOpen
  \bibfield  {author} {\bibinfo {author} {\bibfnamefont {D.~C.}\ \bibnamefont
  {Elton}},\ }\bibfield  {title} {\bibinfo {title} {Stretched exponential
  relaxation},\ }\href@noop {} {\bibfield  {journal} {\bibinfo  {journal}
  {arXiv preprint arXiv:1808.00881}\ } (\bibinfo {year} {2018})}\BibitemShut
  {NoStop}%
\bibitem [{\citenamefont {Johnston}(2006)}]{johnston2006stretched}%
  \BibitemOpen
  \bibfield  {author} {\bibinfo {author} {\bibfnamefont {D.}~\bibnamefont
  {Johnston}},\ }\bibfield  {title} {\bibinfo {title} {Stretched exponential
  relaxation arising from a continuous sum of exponential decays},\ }\href@noop
  {} {\bibfield  {journal} {\bibinfo  {journal} {Physical Review B}\ }\textbf
  {\bibinfo {volume} {74}},\ \bibinfo {pages} {184430} (\bibinfo {year}
  {2006})}\BibitemShut {NoStop}%
\bibitem [{\citenamefont {B{\l}awzdziewicz}\ \emph {et~al.}(2002)\citenamefont
  {B{\l}awzdziewicz}, \citenamefont {Cristini},\ and\ \citenamefont
  {Loewenberg}}]{blawzdziewicz2002critical}%
  \BibitemOpen
  \bibfield  {author} {\bibinfo {author} {\bibfnamefont {J.}~\bibnamefont
  {B{\l}awzdziewicz}}, \bibinfo {author} {\bibfnamefont {V.}~\bibnamefont
  {Cristini}},\ and\ \bibinfo {author} {\bibfnamefont {M.}~\bibnamefont
  {Loewenberg}},\ }\bibfield  {title} {\bibinfo {title} {Critical behavior of
  drops in linear flows. i. phenomenological theory for drop dynamics near
  critical stationary states},\ }\href@noop {} {\bibfield  {journal} {\bibinfo
  {journal} {Physics of Fluids}\ }\textbf {\bibinfo {volume} {14}},\ \bibinfo
  {pages} {2709} (\bibinfo {year} {2002})}\BibitemShut {NoStop}%
\bibitem [{\citenamefont {Windhab}\ \emph {et~al.}(2005)\citenamefont
  {Windhab}, \citenamefont {Dressler}, \citenamefont {Feigl}, \citenamefont
  {Fischer},\ and\ \citenamefont {Megias-Alguacil}}]{windhab2005emulsion}%
  \BibitemOpen
  \bibfield  {author} {\bibinfo {author} {\bibfnamefont {E.~J.}\ \bibnamefont
  {Windhab}}, \bibinfo {author} {\bibfnamefont {M.}~\bibnamefont {Dressler}},
  \bibinfo {author} {\bibfnamefont {K.}~\bibnamefont {Feigl}}, \bibinfo
  {author} {\bibfnamefont {P.}~\bibnamefont {Fischer}},\ and\ \bibinfo {author}
  {\bibfnamefont {D.}~\bibnamefont {Megias-Alguacil}},\ }\bibfield  {title}
  {\bibinfo {title} {Emulsion processing—from single-drop deformation to
  design of complex processes and products},\ }\href@noop {} {\bibfield
  {journal} {\bibinfo  {journal} {Chemical Engineering Science}\ }\textbf
  {\bibinfo {volume} {60}},\ \bibinfo {pages} {2101} (\bibinfo {year}
  {2005})}\BibitemShut {NoStop}%
\bibitem [{\citenamefont {Gordon}\ and\ \citenamefont
  {Schowalter}(1972)}]{gordon1972anisotropic}%
  \BibitemOpen
  \bibfield  {author} {\bibinfo {author} {\bibfnamefont {R.}~\bibnamefont
  {Gordon}}\ and\ \bibinfo {author} {\bibfnamefont {W.}~\bibnamefont
  {Schowalter}},\ }\bibfield  {title} {\bibinfo {title} {Anisotropic fluid
  theory: a different approach to the dumbbell theory of dilute polymer
  solutions},\ }\href@noop {} {\bibfield  {journal} {\bibinfo  {journal}
  {Transactions of the Society of Rheology}\ }\textbf {\bibinfo {volume}
  {16}},\ \bibinfo {pages} {79} (\bibinfo {year} {1972})}\BibitemShut {NoStop}%
\bibitem [{\citenamefont {Elemans}\ \emph {et~al.}(1993)\citenamefont
  {Elemans}, \citenamefont {Bos}, \citenamefont {Janssen},\ and\ \citenamefont
  {Meijer}}]{elemans1993transient}%
  \BibitemOpen
  \bibfield  {author} {\bibinfo {author} {\bibfnamefont {P.}~\bibnamefont
  {Elemans}}, \bibinfo {author} {\bibfnamefont {H.}~\bibnamefont {Bos}},
  \bibinfo {author} {\bibfnamefont {J.}~\bibnamefont {Janssen}},\ and\ \bibinfo
  {author} {\bibfnamefont {H.}~\bibnamefont {Meijer}},\ }\bibfield  {title}
  {\bibinfo {title} {Transient phenomena in dispersive mixing},\ }\href@noop {}
  {\bibfield  {journal} {\bibinfo  {journal} {Chemical engineering science}\
  }\textbf {\bibinfo {volume} {48}},\ \bibinfo {pages} {267} (\bibinfo {year}
  {1993})}\BibitemShut {NoStop}%
\bibitem [{\citenamefont {Meijer}\ \emph {et~al.}(1994)\citenamefont {Meijer},
  \citenamefont {Janssen},\ and\ \citenamefont {Anderson}}]{meijer1994mixing}%
  \BibitemOpen
  \bibfield  {author} {\bibinfo {author} {\bibfnamefont {H.~E.}\ \bibnamefont
  {Meijer}}, \bibinfo {author} {\bibfnamefont {J.~M.}\ \bibnamefont
  {Janssen}},\ and\ \bibinfo {author} {\bibfnamefont {P.~D.}\ \bibnamefont
  {Anderson}},\ }\href@noop {} {\emph {\bibinfo {title} {Mixing of immiscible
  liquids}}}\ (\bibinfo  {publisher} {Hanser, New York},\ \bibinfo {year}
  {1994})\ pp.\ \bibinfo {pages} {85--148}\BibitemShut {NoStop}%
\bibitem [{\citenamefont {Biferale}\ \emph {et~al.}(2014)\citenamefont
  {Biferale}, \citenamefont {Meneveau},\ and\ \citenamefont
  {Verzicco}}]{biferale2014deformation}%
  \BibitemOpen
  \bibfield  {author} {\bibinfo {author} {\bibfnamefont {L.}~\bibnamefont
  {Biferale}}, \bibinfo {author} {\bibfnamefont {C.}~\bibnamefont {Meneveau}},\
  and\ \bibinfo {author} {\bibfnamefont {R.}~\bibnamefont {Verzicco}},\
  }\bibfield  {title} {\bibinfo {title} {Deformation statistics of
  sub-kolmogorov-scale ellipsoidal neutrally buoyant drops in isotropic
  turbulence},\ }\href@noop {} {\bibfield  {journal} {\bibinfo  {journal}
  {Journal of Fluid Mechanics}\ }\textbf {\bibinfo {volume} {754}},\ \bibinfo
  {pages} {184} (\bibinfo {year} {2014})}\BibitemShut {NoStop}%
\bibitem [{\citenamefont {Elghobashi}(2019)}]{elghobashi2019direct}%
  \BibitemOpen
  \bibfield  {author} {\bibinfo {author} {\bibfnamefont {S.}~\bibnamefont
  {Elghobashi}},\ }\bibfield  {title} {\bibinfo {title} {Direct numerical
  simulation of turbulent flows laden with droplets or bubbles},\ }\href@noop
  {} {\bibfield  {journal} {\bibinfo  {journal} {Annual Review of Fluid
  Mechanics}\ }\textbf {\bibinfo {volume} {51}},\ \bibinfo {pages} {217}
  (\bibinfo {year} {2019})}\BibitemShut {NoStop}%
\bibitem [{\citenamefont {Spandan}\ \emph {et~al.}(2016)\citenamefont
  {Spandan}, \citenamefont {Lohse},\ and\ \citenamefont
  {Verzicco}}]{spandan2016deformation}%
  \BibitemOpen
  \bibfield  {author} {\bibinfo {author} {\bibfnamefont {V.}~\bibnamefont
  {Spandan}}, \bibinfo {author} {\bibfnamefont {D.}~\bibnamefont {Lohse}},\
  and\ \bibinfo {author} {\bibfnamefont {R.}~\bibnamefont {Verzicco}},\
  }\bibfield  {title} {\bibinfo {title} {Deformation and orientation statistics
  of neutrally buoyant sub-kolmogorov ellipsoidal droplets in turbulent
  taylor--couette flow},\ }\href@noop {} {\bibfield  {journal} {\bibinfo
  {journal} {Journal of Fluid Mechanics}\ }\textbf {\bibinfo {volume} {809}},\
  \bibinfo {pages} {480} (\bibinfo {year} {2016})}\BibitemShut {NoStop}%
\end{thebibliography}%

\end{document}